\documentclass{article}

\usepackage{arxiv}

\usepackage{psfrag,epsf}
\usepackage{enumerate}
\usepackage[utf8]{inputenc} 
\usepackage[T1]{fontenc}    
\usepackage[english]{babel}		
\usepackage{lmodern}
\usepackage{hyperref}       
\usepackage{url}            
\usepackage{booktabs}       
\usepackage{amsfonts}       
\usepackage{nicefrac}       
\usepackage{microtype}      
\usepackage{float}          
\usepackage{bm}    
\usepackage{bbm}         
\usepackage{amssymb}        
\usepackage{amsmath}        
\usepackage{graphicx}       
\usepackage{comment}
\usepackage[font=footnotesize,labelfont=bf]{caption}
\usepackage{subcaption}
\usepackage{tikz}           
\usepackage[shortlabels]{enumitem}
\usepackage{array}          
\usepackage{lscape}         
\usepackage{dirtytalk}      
\usepackage{mathtools}      
\usepackage{tabularx}  
\usepackage{siunitx}
\usepackage{etoolbox}
\usepackage{cleveref}             
\usepackage{doi}
\usepackage{natbib}
\usepackage{adjustbox}
\usepackage{multirow}
\usepackage{multicol}
\usepackage{pdflscape}
\usepackage{makecell}
\usepackage{pifont}
\newcommand{\cmark}{\ding{51}}

\graphicspath{{./assets/}}

\title{Semi-structured multi-state delinquency model for mortgage default}

\date{}

\author{
  Victor Medina-Olivares\thanks{Victor.Medina@ed.ac.uk}\\
  Business School, \\
  The University of Edinburgh, UK.
	\And
	Wangzhen Xia \\
	School of Management,\\
  Technische Universität München, Germany.
	\And
	Stefan Lessmann \\
  School of Business and Economics,\\ 
  Humboldt-Universität zu Berlin, Germany.
  \And
	Nadja Klein \\
  Scientific Computing Center,\\ 
  Karlsruhe Institute of Technology, Germany. 
}

\renewcommand{\shorttitle}{Semi-structured multi-state delinquency model}

\hypersetup{
pdftitle={Semi-structured multi-state delinquency model},
pdfsubject={stat.ML, cs.LG},
pdfauthor={Victor ~Medina-Olivares},
pdfkeywords={Credit risk, Deep learning, Interpretability, Loan-level data, Transition probabilities, Survival analysis},
}

\begin{document}

\maketitle

\begin{abstract}
We propose a semi-structured discrete-time multi-state model to analyse mortgage delinquency transitions. This model combines an easy-to-understand structured additive predictor, which includes linear effects and smooth functions of time and covariates, with a flexible neural network component that captures complex nonlinearities and higher-order interactions. To ensure identifiability when covariates are present in both components, we orthogonalise the unstructured part relative to the structured design. For discrete-time competing transitions, we derive exact transformations that map binary logistic models to valid competing transition probabilities, avoiding the need for continuous-time approximations. In simulations, our framework effectively recovers structured baseline and covariate effects while using the neural component to detect interaction patterns. We demonstrate the method using the Freddie Mac Single-Family Loan-Level Dataset, employing an out-of-time test design. Compared with a structured generalised additive benchmark, the semi-structured model provides modest but consistent gains in discrimination across the earliest prediction spans, while maintaining similar Brier scores. Adding macroeconomic indicators provides limited incremental benefit in this out-of-time evaluation and does not materially change the estimated borrower-, loan-, or duration-driven effects. Overall, semi-structured multi-state modelling offers a practical compromise between transparent effect estimates and flexible pattern learning, with potential applications beyond credit-transition forecasting.
\end{abstract}


\section{Introduction} \label{sec:intro}
Multi-state models extend survival analysis by modelling transitions among multiple states over time, with transition probabilities that may depend on covariates and time \citep{hougaard1999multi,cook2018multistate}. This framework captures both the timing and sequence of events and is particularly well-suited to credit risk, where borrower behaviour evolves through states such as timely payment, early delinquency, late delinquency, and default. In contrast to two-state models that classify borrowers simply as ``default'' or ``non-default'' \citep{bellotti2013forecasting}, multi-state approaches offer a richer description of the dynamics of credit portfolios. In particular, this additional structure allows us to separate entry into delinquency, escalation across delinquency states, and cure back to current. This enables state-contingent forecasting (e.g., expected delinquency-bucket volumes over time) and supports collections and provisioning decisions beyond aggregate default probabilities. We next summarise how these ideas have been developed in this context, in particular, corporate rating-migration models and retail credit applications.

In corporate credit risk, multi-state models have long been used to analyse rating migrations, where firms move among discrete rating categories over time. The literature is broadly divided into continuous-time and discrete-time frameworks. The former allows transitions to occur at any instant, whereas the latter restricts transitions to fixed observation intervals. Continuous-time frameworks include specifications without covariates, such as standard continuous-time Markov chain models and continuous-time hidden Markov models \citep{Jafry2004, Christensen2004, Bladt2009}. Extensions with covariates typically use multiplicative (proportional) hazards to incorporate firm-specific or macroeconomic drivers \citep{Lando2002,Koopman2009}. In practice, however, these models often accommodate only a limited set of covariates and impose relatively simple functional forms. Although continuous-time specifications are flexible and theoretically appealing, they may still require restrictive assumptions or computational approximations. In many credit datasets, ratings and covariates are observed only at regular reporting dates, such as monthly or quarterly intervals. In such settings, a discrete-time formulation is often the more natural modelling choice. Discrete-time approaches include, for example, discrete-time hidden Markov models \citep{Korolkiewicz2008} and dynamic ordered logit models \citep{Creal2014}.

Retail credit raises similar questions, but it typically involves higher-frequency behavioural states, such as delinquency buckets, together with richer borrower-level information. This makes multi-state models an appealing framework for analysing the evolution of loan states, while also increasing the need for flexible specifications that can capture nonlinear and interaction effects. As in corporate rating-transition modelling, multiplicative hazard models are widely used to analyse transitions across loan states such as delinquency and prepayment \citep{Leow2014, Kelly2016, Bocchio2023}. Another common approach combines a Markov chain with a multinomial logit model \citep{Nystrom2006, Grimshaw2011}.

Because these models follow the evolution of loan states over time, it is also natural to consider dependence among repeated observations on the same loan. One way to address this issue is through unobserved heterogeneity. For example, \citet{Djeundje2018} augment a marginal Bernoulli process with random effects (i.e., frailty) to model transitions of credit card loans across delinquency states. They find that including frailty does not improve predictive accuracy, although it reduces the apparent significance of the observed covariates. By contrast, \citet{Bocchio2023} use variance-corrected methods to account for within-subject correlation induced by repeated events. The need for flexible functional forms is also evident in work such as \citet{Sadhwani2020}, who use deep neural networks to incorporate nonlinear and interaction effects in the multi-state modelling of residential mortgage loans.

Overall, these studies reflect a growing emphasis on flexible representations of credit dynamics that can capture, among other things, nonlinear effects and higher-order interactions. This flexibility must, however, be balanced against the continuing requirement for interpretability in regulated settings, where lenders must be able to interpret and explain model-driven decisions. At the same time, more expressive models provide a natural route for future extensions that incorporate non-traditional data sources (e.g., transactional cash-flow data from open-banking settings, as well as other non-structured information) that are increasingly available and may further improve predictive performance.

\paragraph*{Our contributions.}
We contribute to the literature in three ways. First, we propose a semi-structured discrete-time multi-state model that combines (i) an interpretable structured additive predictor (linear terms and splines) with (ii) an unstructured neural-network component designed to capture complex nonlinearities and high-dimensional interactions. To ensure identifiability and prevent the neural component from absorbing the structured signal, we apply an orthogonalisation step that separates the structured and unstructured components \citep{rugamer2024semi}. Second, we derive exact probability transformations that map a collection of non-competing binary logistic transition models to coherent competing transition probabilities in discrete time. This yields an explicit discrete-time probability model aligned with the reporting frequency of credit data and avoids continuous-time approximations commonly used in related credit and actuarial work \citep{Djeundje2018,Luptakova2014}. Third, we provide an empirical evaluation on large-scale US mortgage data with out-of-time testing, benchmarking against a generalised additive model without the neural component, and quantifying the marginal contribution of macroeconomic variables. The results indicate that, for this dataset, the proposed model improves predictive performance---particularly discrimination---while retaining interpretability for the structured effects. Beyond the present application, the semi-structured design is compatible with future extensions that incorporate unstructured inputs when available. One example is biostatistics, where survival analyses are ubiquitous and imaging data, such as MRI scans, are increasingly available.

Table~\ref{tab:refs} provides an overview of representative multi-state studies in credit risk, including those discussed above, and highlights a recurring trade-off between model flexibility and interpretability. It also positions our approach relative to prior works. In the table, ``\textit{Application}'' shows the area to which a model is applied. ``\textit{Time}'' specifies whether the study is conducted within a discrete- or continuous-time framework. ``\textit{Interpretable}'' refers to whether key drivers can be communicated through explicit parameters in the model (e.g., log-odds or hazard ratios and smooth effects) rather than only through a black-box mapping. ``\textit{Nonlinear effects}'' indicates whether the model can capture the nonlinear relationship between covariates and the response variable beyond simple linear terms (e.g., via splines, time-varying coefficients, or flexible machine-learning components). ``\textit{Interactions}'' indicates whether higher-order covariate combinations are accounted for (e.g., through explicit interaction terms, latent-factor structure, or neural components). ``\textit{Time-dep. baseline}'' shows whether duration dependence is considered. Finally, ``\textit{Heterogeneity}'' summarises whether the model takes unobserved loan-level variation into account.

The remainder of the paper is organised as follows. Section~\ref{sec:data} describes the Freddie Mac mortgage panel and the delinquency-state construction. Section~\ref{sec:method} presents the modelling framework, Section~\ref{sec:sim} reports simulations, Section~\ref{sec:emp_res} provides the empirical evaluation, and Section~\ref{sec:concl} concludes.

\begin{table}
    \centering
    \begin{adjustbox}{max width=\textwidth}
    \begin{tabular}{lllllccccc}
    \toprule
        \multicolumn{1}{c}{\textbf{Reference}} &
        \multicolumn{1}{c}{\textbf{Application}} &
        \textbf{Time} &
        \textbf{\# Units} &
        \multicolumn{1}{c}{\textbf{Approach}} &
        \textbf{Interpretable} &
        \textbf{Nonlinear effects} &
        \textbf{Interactions} &
        \textbf{Time-dep. baseline} &
        \textbf{Heterogeneity} \\
    \midrule
        \citet{Bangia2002} & Corporate rating & $N_{+}$ & \num{7328} & Cohort estimation &  &  &  &  &  \\
        \citet{Lando2002} & Corporate rating & $R_{+}$ & \num{6659} & Multiplicative hazard model & \cmark &  &  & \cmark &  \\
        \citet{Christensen2004} & Corporate rating & $R_{+}$ & \num{3446} & Hidden Markov chain model &  &  &  &  &  \\
        \multirow{2}{*}{\citet{Jafry2004}} & \multirow{2}{*}{Corporate rating}  & $N_{+}$ & \multirow{2}{*}{\num{6776}} & Cohort estimation &  &  &  &  &  \\
         & & $R_{+}$ & & Markov chain model &  &  &  &  &  \\
        \citet{Nystrom2006} & Loan states & $N_{+}$ & \num{160000} & \makecell[l]{Markov chain model combined with \\multinomial Logit} & \cmark &  &  &  &  \\
        \citet{Koopman2008} & Corporate rating & $R_{+}$ & $\sim$ \num{7000} & Multiplicative hazard model & \cmark &  &  & \cmark &  \\
        \citet{Korolkiewicz2008} & Corporate rating & $N_{+}$ & \num{1301} & Hidden Markov chain model & & & & & \\
        \citet{Bladt2009} & Corporate rating  & $R_{+}$ & \num{2749} & Markov chain model &  &  &  &  &  \\
        \citet{Koopman2009} & Corporate rating  & $R_{+}$ & / & Multiplicative hazard model & \cmark &  &  & \cmark &  \\
        \citet{Grimshaw2011} & Subprime mortgage delinquency & $N_{+}$ & \num{97124} & \makecell[l]{Markov chain model combined with \\multinomial Logit} & \cmark &  &  & \cmark &  \\
        \multirow{2}{*}{\citet{Schechtman2013}} & \multirow{2}{*}{Consumer credit delinquency} & $N_{+}$ & \multirow{2}{*}{/} & Cohort estimation &  &  &  &  &  \\
         &  & $R_{+}$ & & Markov chain model &  &  &  &  &  \\
        \citet{Creal2014} & Corporate rating  & $N_{+}$ & \num{7505} & Dynamic ordered Logit model & \cmark &  &  &  &  \\
        \citet{Leow2014} & Credit card loan delinquency & $N_{+}$ & $\sim$ \num{49000} & Multiplicative hazard model & \cmark &  &  & \cmark &  \\
        \citet{Kelly2016} & Residential mortgage delinquency & $R_{+}$ & \num{64669} & Multiplicative hazard model & \cmark &  &  & \cmark &  \\
        \citet{Djeundje2018} & Credit card loan delinquency & $N_{+}$ & $\sim$ \num{35000} & Marginal Bernoulli processes & \cmark &  &  & \cmark & \cmark \\
        \citet{Sadhwani2020} & Residential mortgage states & $N_{+}$ & $\sim$ 120M & Ensemble of NNets &  & \cmark & \cmark & \cmark &  \\
        \citet{Bocchio2023} & Residential mortgage delinquency & $R_{+}$ & \num{67827} & Multiplicative hazard model & \cmark &  &  & \cmark & \cmark \\
        \citet{djeundje2025devil} & Residential mortgage states & $N_{+}$ & $\sim$ \num{100000} & Marginal Bernoulli processes & \cmark &  &  & \cmark &  \\
        \citet{Koffi2025} & Microloan delinquency & $N_{+}$ & \num{1716} & Marginal Bernoulli processes & \cmark &  &  & \cmark & \cmark \\
        \textbf{Our study} & Residential mortgage delinquency & $N_{+}$ & $\sim$ \num{150000} & Semi-structured regression & \textbf{\cmark} & \textbf{\cmark} & \textbf{\cmark} & \textbf{\cmark} & \textbf{\cmark} \\
    \bottomrule
    \end{tabular}
    \end{adjustbox}
    \caption{References in credit risk literature using multi-state approaches.
    }
    \label{tab:refs}
\end{table}

\section{US Mortgage Data} \label{sec:data}
For the empirical analysis, we use the Freddie Mac Single-Family Loan-Level Dataset for cohorts from 2016 to 2019. The training set includes approximately \num{100000} loans with first payment dates between March 2016 and February 2018, while the test set has around \num{50000} loans with first payment dates between March 2018 and February 2019. This design ensures both out-of-sample and out-of-time testing. The window also provides at least three years of payment history per loan, as the latest observation in our extract is December 2022. The training and test sets contain \num{4286349} and \num{1435505} monthly observations, respectively.

In addition to loan duration, we control for origination characteristics, a binary indicator for the COVID-19 period (March--December 2020), and several macroeconomic variables. The macro indicators are the S\&P~500 index, the Consumer Confidence Index (CCI), the 30-year mortgage rate (MORT30), the Industrial Production Index (IPI), the Consumer Price Index (CPI), and the unemployment rate (UNRT). All indicators are lagged by six months and normalised to their January 2000 levels (or February 2000 where January is unavailable).

Given the granularity of the data and the business context of modelling state transitions, we define four delinquency states:
\begin{itemize}[leftmargin=2cm]
  \item[\textbf{State 0}]: The borrower is fewer than 30 days past due.
  \item[\textbf{State 1}]: The borrower is 30-59 days past due.
  \item[\textbf{State 2}]: The borrower is 60-89 days past due.
  \item[\textbf{State 3}]: The borrower is 90 days or more past due (default).
\end{itemize}

\begin{figure}
	\centering
    \includegraphics[width=0.85\textwidth]{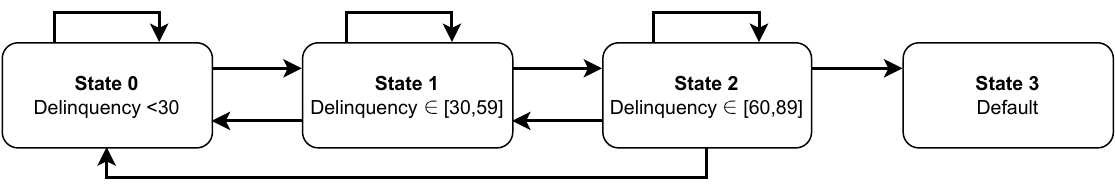}
	\caption{Delinquency states and permissible transitions over a one-month period.}
	\label{fig:trans}
\end{figure}

The permissible month-to-month transitions are shown in Figure~\ref{fig:trans}. For example, a loan in state $0$ can either remain in state $0$ or move to state $1$ over a one-month period. A loan in state $1$ can move to state $2$, return to state $0$, or remain in state $1$. A loan in state $2$ can transition to state $3$, return to state $1$, or return to state $0$. State $3$ is absorbing.

\begin{figure}
	\centering
    \includegraphics[width=0.85\textwidth]{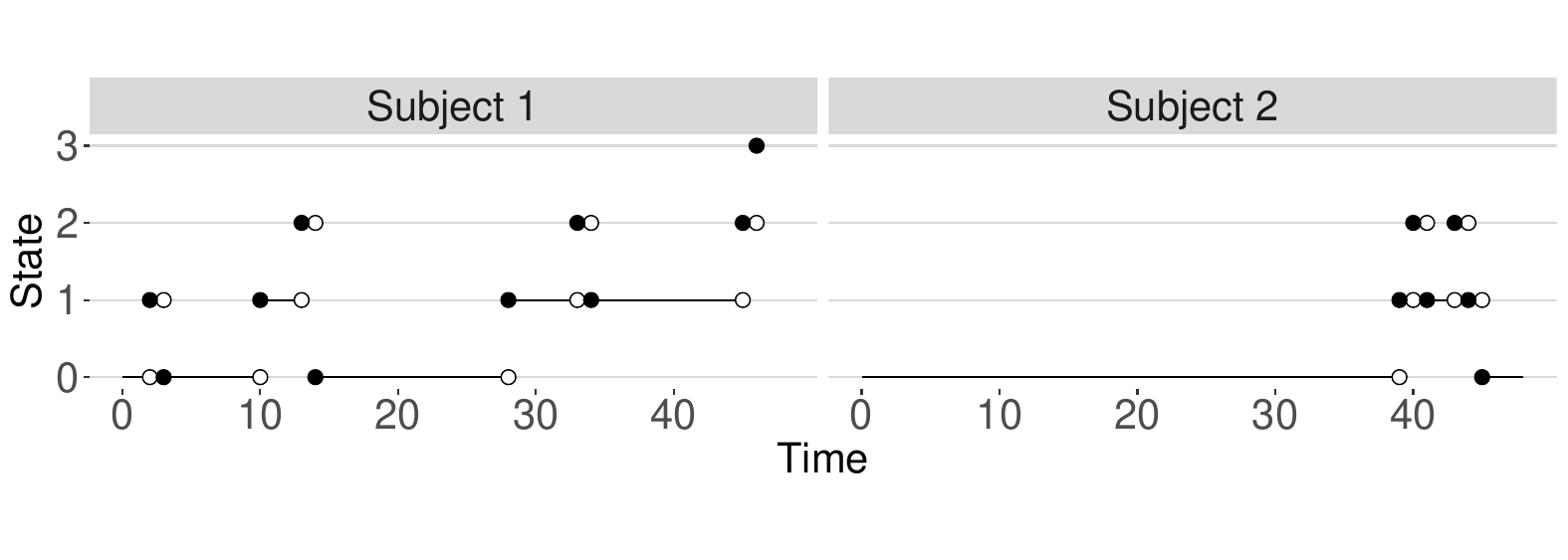}
    \caption{Realisation of a process path for two randomly selected subjects in the training set.}
  \label{fig:path}
\end{figure}

Figure \ref{fig:path} shows the realisation of two randomly selected individuals for whom at least one transition occurs. Both start in state 0 at origination. Subject 1 transitions to state 1, returns to 0, then reverts back to state 1 and moves up to state 2, repays the overdue months, and returns to 0; after several such episodes, the subject eventually defaults. Subject 2, after remaining in state 0 for a long time, also shows episodes of increased delinquency, but eventually returns to state 0 and remains there until the end of the study period.

\begin{figure}
	\centering
    \includegraphics[width=0.85\textwidth]{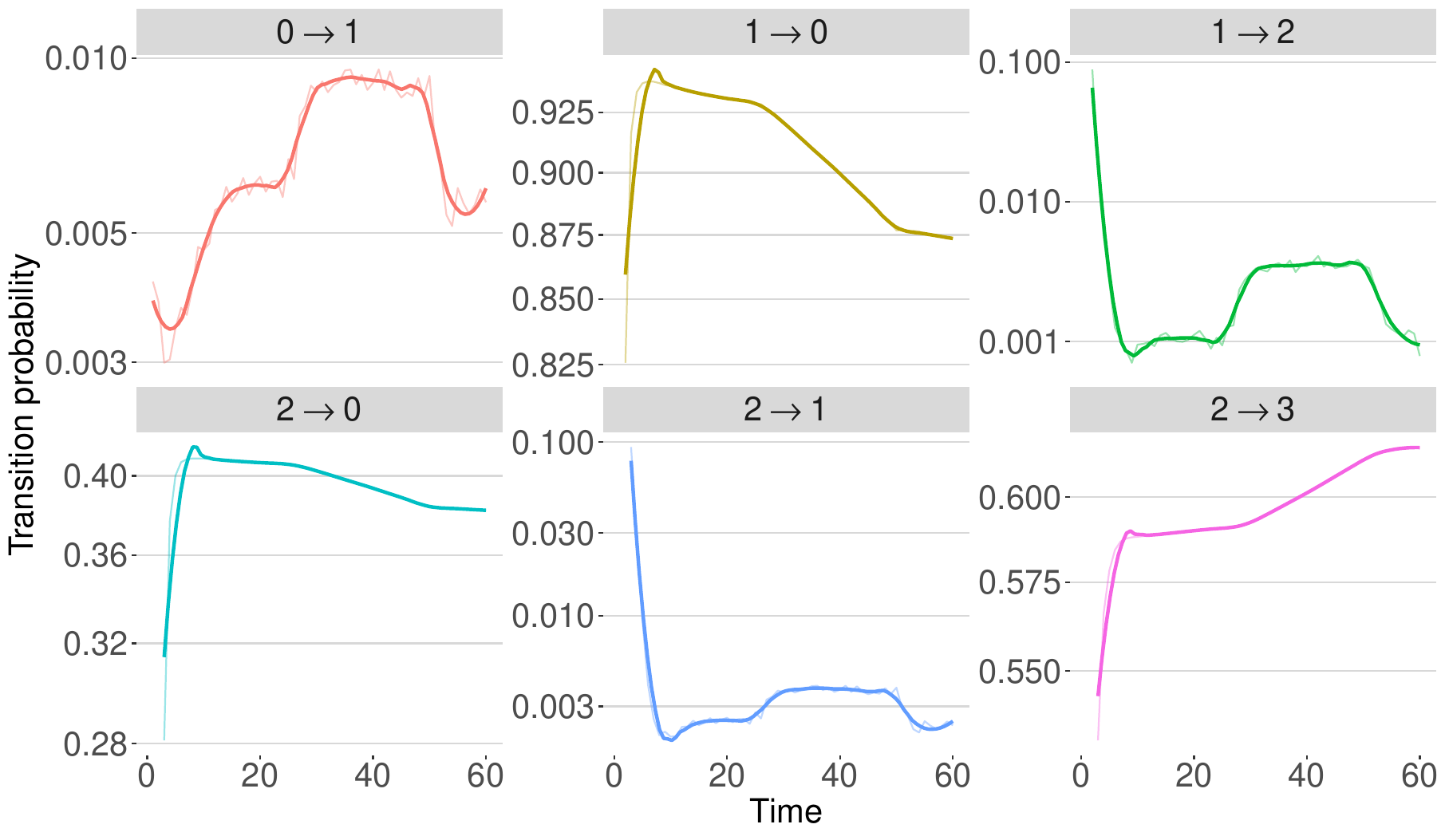}
  \caption{Empirical transition probabilities (Aalen--Johansen estimator) for the U.S.\ mortgage loan dataset. The y-axis is on a log scale for readability. Solid lines show the raw estimates, and the overlaid smooth curves (LOESS) highlight the underlying time pattern.}
  \label{fig:aalen}
\end{figure}

In Figure \ref{fig:aalen}, we present the empirical transition matrix, i.e., the Aalen-Johansen estimator \citep{allignol2011empirical}, for the permissible transitions (cf.\ Figure \ref{fig:trans}). The interpretation is straightforward. For example, the empirical probability that an individual who is in state~0 at the start of the study has transitioned to state~1 by month~40 is about 1\%. Likewise, the probability of being in default (state~3) by month~40, conditional on being in state~2 at the beginning, is around 60\%. We also observe a clear increase in transitions---particularly $0 \rightarrow 1$, $1 \rightarrow 2$, and $2 \rightarrow 1$---which coincides with the COVID-19 period.

The upper panel of Table \ref{tab:events} reports the number of loans undergoing different transition types in the training set. Most loans in state $0$ at the start of a month remain in state $0$ by month-end. Approximately half of loans in state $1$ return to state $0$, while more than one-third transition to state $2$. For loans that reach state $2$, the probability of default exceeds 60\%. Some loans recover from state $2$ to state $0$ (17.4\%) or state $1$ (7.0\%). The lower panel provides analogous information for the test set. The empirical distribution of transitions is similar across the two sets.
\begin{table}
    \centering
    \begin{tabular}{rrrrrrr}
        \toprule
            \multicolumn{7}{l}{\textit{Training Set}} \\
             & & \multicolumn{5}{c}{\textbf{To state}} \\
             & & \multicolumn{1}{c}{0} & \multicolumn{1}{c}{1} & \multicolumn{1}{c}{2} & \multicolumn{1}{c}{3} & Total \\
            \multirow{6}{*}{\textbf{From state}} & \multirow{2}{*}{0} & 95,866 & 12,476 & - & - & 108,342 \\
             & & (88.5\%) & (11.5\%) & - & - & (100\%) \\
             & \multirow{2}{*}{1} & 8,245 & 2,674 & 5,860 & - & 16,779 \\
             & & (49.1\%) & (15.9\%) & (34.9\%) & - & (100\%) \\
             & \multirow{2}{*}{2} & 1,243 & 503 & 863 & 4,530 & 7,139 \\
             & & (17.4\%) & (7.0\%) & (12.1\%) & (63.5\%) & (100\%) \\
        \midrule
            \multicolumn{7}{l}{\textit{Test Set}} \\
            & & \multicolumn{5}{c}{\textbf{To state}} \\
            & & \multicolumn{1}{c}{0} & \multicolumn{1}{c}{1} & \multicolumn{1}{c}{2} & \multicolumn{1}{c}{3} & Total \\
            \multirow{6}{*}{\textbf{From state}} & \multirow{2}{*}{0} & 47,178 & 5,865 & - & - & 53,043 \\
             & & (88.9\%) & (11.1\%) & - & - & (100\%) \\ 
             & \multirow{2}{*}{1} & 3,565 & 1,174 & 3,009 & - & 7,748 \\ 
             & & (46.0\%) & (15.2\%) & (38.8\%) & - & (100\%) \\ 
             & \multirow{2}{*}{2} & 587 & 227 & 401 & 2,419 & 3,634 \\
             & & (16.2\%) & (6.2\%) & (11.0\%) & (66.6\%) & (100\%) \\
        \bottomrule
    \end{tabular}
    \caption{Frequency of loans for different transition types.}
    \label{tab:events}
\end{table}

\begin{figure}
    \centering
    \includegraphics[width=0.8\linewidth]{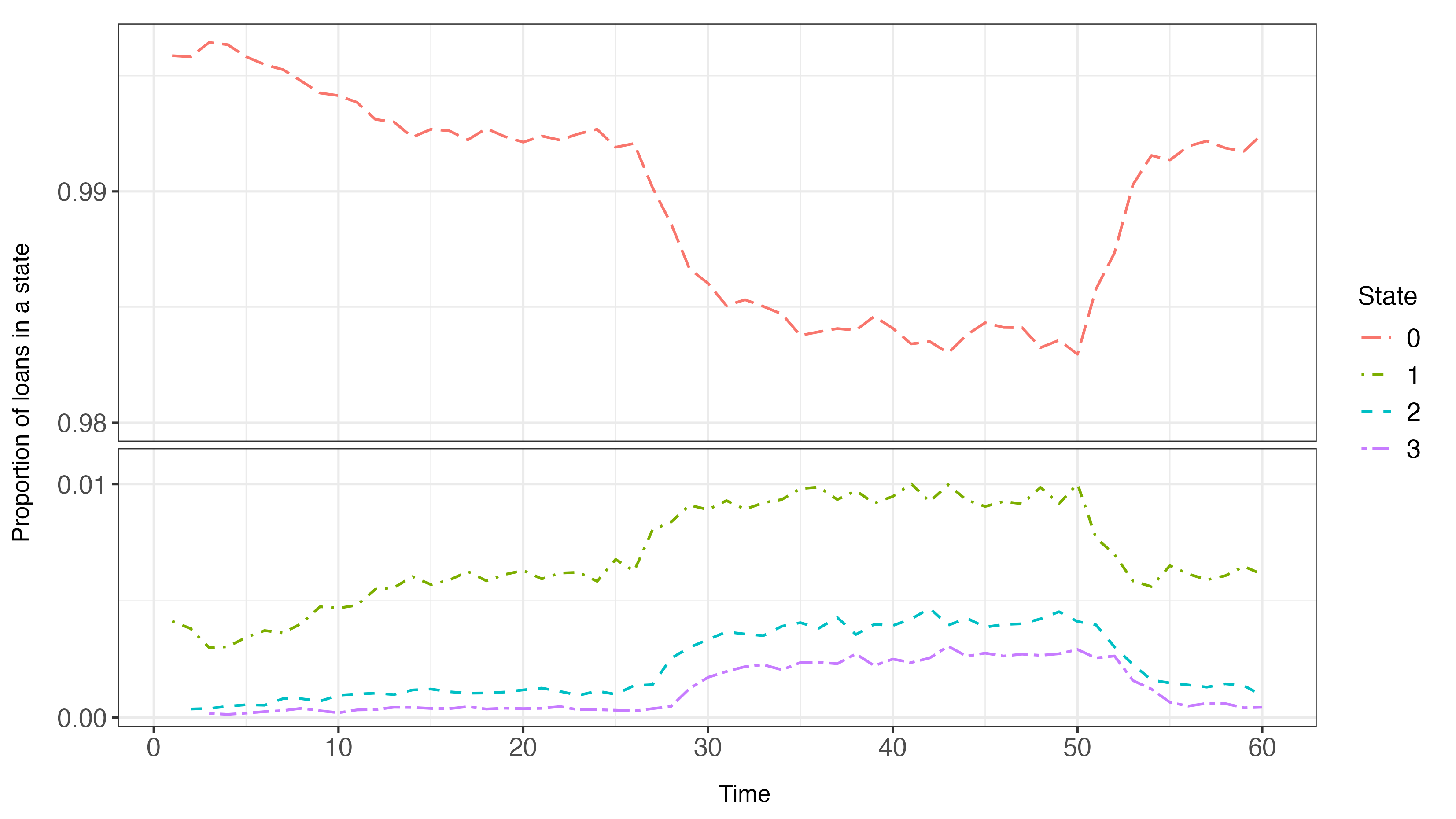}
    \caption{Proportion of loans in training set in each state at month-end since origination.}
    \label{fig:to_state_train}
\end{figure}

Figure \ref{fig:to_state_train} shows the proportion of loans in each state during the first 60 months since origination, irrespective of the subsequent transition. Most loans remain in state $0$. There is a gradual decline in the share of loans in state $0$ over the first 50 months, which accelerates during the pandemic period. The empirical frequency of loans in state $1$ is nearly three times that in state $3$ (default), and the share in state $2$ is similar to that in default.

Table \ref{tab:sum_stat_num_tv} summarises the numerical variables in the training set before standardisation. Credit scores range from \num{386} to \num{832}, with a mean of \num{747.81} and a standard deviation of \num{46.72}. The maximum interest rate is more than twice the minimum and around 1.5 times the mean or median. The loan-to-value ratio has a mean of \num{73.85}\% and a maximum of \num{339}\%. Most mortgages originated since March 2016 do not have mortgage insurance cover. The median loan term is 30 years. As expected, there is substantial variation in the unpaid principal balance (UPB).
\begin{table}
  \centering
  \begin{tabular}{lrrrrr}
      \toprule
          \multicolumn{1}{c}{\textbf{Variable}} & \multicolumn{1}{c}{\textbf{Min.}} & \multicolumn{1}{c}{\textbf{Mean}} & \multicolumn{1}{c}{\textbf{Median}} & \multicolumn{1}{c}{\textbf{Max.}} & \multicolumn{1}{c}{\textbf{SD.}} \\
      \midrule
          \textit{Credit Score} & 386.00 & 747.81 & 756.00 & 832.00 & 46.72 \\
          \textit{Original Interest Rate (\%)} & 2.25 & 3.96 & 3.99 & 6.12 & 0.50 \\
          \textit{Original Loan-to-Value Ratio (\%)} & 6.00 & 73.85 & 79.00 & 339.00 & 17.34 \\
          \textit{Mortgage Insurance (\%)} & 0.00 & 6.60 & 0.00 & 40.00 & 11.50 \\
          \textit{Original Loan Term (months)} & 96.00 & 318.42 & 360.00 & 360.00 & 74.08 \\
          \textit{Original UPB (\$)} & 14,000.00 & 225,947.27 & 204,000.00 & 985,000.00 & 117,011.18 \\
      \bottomrule
  \end{tabular}
  \caption{Summary statistics for numerical variables in the training set. Further variable descriptions are available in the \href{https://www.freddiemac.com/research/datasets/sf-loanlevel-dataset}{General User Guide}  for the Freddie Mac Single-Family Loan-Level Dataset. Numerical variables are normalised prior to modelling.}
  \label{tab:sum_stat_num_tv}
\end{table}

Table \ref{tab:sum_stat_cat_tv} reports the cardinalities of the categorical variables. The debt-to-income (DTI) ratio is continuous in the raw data. However, we use a discretised DTI because the dataset takes only a limited set of fixed values and does not disclose DTI when it exceeds 65\% or qualifies for the Home Affordable Refinance Programme (HARP)\footnote{HARP is a programme designed to help borrowers who cannot refinance their mortgages due to depreciated property values. Further details are available in Freddie Mac's HARP documentation.}. The cut-points for discretising DTI are 20\%, 30\%, and 40\%, with missing DTI assigned to a separate category. To facilitate model training, we preprocess high-cardinality categorical covariates---Seller Name, Servicer Name, and US state/territory---using a weight-of-evidence (WoE) encoding \citep{Moeyersoms15WOE}. Tables~\ref{tab:sum_stat_num_t} and \ref{tab:sum_stat_cat_t} in Appendix~\ref{app:sum_stat} provide the corresponding summary statistics for the test set.

\begin{table}
    \centering
    \begin{tabular}{l r}
        \toprule
            \textbf{Variable(s)} & \textbf{Cardinality} \\
        \midrule
            \textit{Number of Borrowers}, \textit{First Time Homebuyer}, \textit{Super Conforming} & 2 \\
            \textit{Channel}, \textit{Loan Purpose}, \textit{Mortgage Insurance Cancellation}, \textit{Occupancy Status} & 3 \\
            \textit{Number of Units} & 4 \\
            \textit{Original Debt-to-Income Ratio}, \textit{Property Type} & 5 \\
            \textit{Seller Name}, \textit{Servicer Name} & 31 \\
            \textit{US state or territory} & 54 \\
        \bottomrule
    \end{tabular}
    \caption{Cardinality of categorical loan characteristics in the training set. Further variable descriptions are available in the \href{https://www.freddiemac.com/research/datasets/sf-loanlevel-dataset}{General User Guide} for the Freddie Mac Single-Family Loan-Level Dataset.}
    \label{tab:sum_stat_cat_tv}
\end{table}

\section{Methodology} \label{sec:method}
This section defines the discrete-time multi-state framework used throughout (Section \ref{subsec:multi-state}) and then introduces our semi-structured predictor, which combines an interpretable structured component with a flexible neural component (Section \ref{subsec:semi-str}). We then discuss identifiability via orthogonalisation (Section \ref{subsec:identifiability}) and the exact transformation from binary logistic transition models to valid competing transition probabilities (Section \ref{subsec:competing}).

\subsection{Discrete-time multi-state model} \label{subsec:multi-state}
Consider a $K$-state process where $Z_i(t)$ denotes the state of subject $i$ at discrete times $t = 0, 1, \dots, T$, with $t$ representing duration (typically time since loan origination) and $T$ the length of the study. The model incorporates a $p \times 1$ vector of covariates $\bm{x}_i(t)$, which may be time-independent (fixed at loan application date $\tau_i$ in our case) or time-dependent (e.g., macroeconomic variables evolving over calendar time). Calendar time can be expressed relative to $\tau_i$ and duration $t$. In our application, macroeconomic variables are often lagged for prediction purposes \citep{bellotti2009credit}. Moreover, as macroeconomic paths do not depend on the payment behaviour of any particular individual, we treat these covariates as exogenous to the transitions \citep[cf.][]{medina2025joint}.

Let $Y_{ik}(t)$ be the indicator variable defined by $Y_{ik}(t)=1$ if $Z_i(t)=k$ and $0$ otherwise, for $k=0,\dots,K-1$. Define $N_{ikl}(t)$ as the indicator of a transition from state $k$ to state $l$ at time $t$, with $N_{ikl}(t)=1$ if $Z_i(t-1)=k$ and $Z_i(t)=l$, and $0$ otherwise. The vector $\bm{N}_{ik}(t)=[N_{ik0}(t),\dots,N_{ik,K-1}(t)]^\top$ thus follows a multinomial distribution \citep{cook2018multistate}.

In the four-state delinquency process (cf.\ Figure~\ref{fig:trans}), we define $\bm{N}_{i0}(t)$, $\bm{N}_{i1}(t)$, $\bm{N}_{i2}(t)$, and $\bm{N}_{i3}(t)$ for each borrower $i$. As noted in Section~\ref{sec:data}, state $k=3$ is absorbing, so that $\bm{N}_{i3}(t)=(0,0,0,1)^\top$ for all $i$.

Let the history of the process of borrower $i$ at time $s$ be $\mathcal{H}_i(s)=\lbrace(Z_i(t),\bm{x}_i(t)),\, t=0,1,\dots,s\rbrace$. Under a first-order Markov assumption and observations up to time $s$, the probability of observing a path of length $s$ for borrower $i$ given the starting state at time $t=0$ and the covariate information up to $t=s-1$ is 
\begin{equation*}
  P\!\left(Z_i(1),\dots,Z_i(s)\mid Z_i(0),\bm{x}_i(0),\dots,\bm{x}_i(s-1)\right)
  = \prod_{t=1}^s P\!\left(Z_i(t)\mid \mathcal{H}_i(t-1)\right)
  = \prod_{t=1}^s P\!\left(Z_i(t)\mid Z_i(t-1),\bm{x}_i(t-1)\right),
\end{equation*}
where 
\begin{equation*}
  P\!\left(Z_i(t)\mid Z_i(t-1),\bm{x}_i(t-1)\right)
  = \prod_{k=0}^{K-1}\left(\prod_{l=0}^{K-1}\pi_{ikl}(t)^{N_{ikl}(t)}\right)^{Y_{ik}(t-1)},
\end{equation*}
with $\pi_{ikl}(t)=P\!\left(Z_i(t)=l \mid Z_i(t-1)=k,\bm{x}_i(t-1)\right)$ the transition probability from state $k$ to $l$ at time $t$ for subject $i$.

For $N$ subjects, each observed up to $T_i=\min\{C_i,T_i^\ast\}$, where $C_i$ is the last observed time (e.g.\, due to study termination or loss to follow-up) and $T_i^\ast$ is the time to the absorbing state (credit default in our case), the likelihood is
\begin{equation}
L \;=\; \prod_{i=1}^N \prod_{t=1}^{T_i} \prod_{k=0}^{K-1}
\left( \prod_{l=0}^{K-1} \pi_{ikl}(t)^{N_{ikl}(t)} \right)^{Y_{ik}(t-1)}.
\label{eq:lik}
\end{equation}

We parameterise the log-odds ratio between $\pi_{ikl}(t)$ and $\pi_{ikk}(t)$ via the predictor $\eta_{ikl}(t)\in\mathbb{R}$,
\begin{equation}
\label{eq:logodds}
\log\!\left(\frac{\pi_{ikl}(t)}{\pi_{ikk}(t)}\right)=\eta_{ikl}(t), \qquad l=0,\dots,K-1,
\end{equation}
which implies the multinomial logits for the transition probabilities, that is, 
\begin{equation*}
\pi_{ikl}(t) = \frac{\exp\{\eta_{ikl}(t)\}}{\sum_{l'=0}^{K-1}\exp\{\eta_{ikl'}(t)\}}, \qquad
\eta_{ikk}(t)=0,\quad \sum_{l=0}^{K-1}\pi_{ikl}(t)=1 , \qquad\mbox{ for all } t.
\end{equation*}
In principle, \eqref{eq:lik}--\eqref{eq:logodds} can be estimated via a multinomial logit specification. In our setting, however, estimation is implemented by fitting separate binary logistic regressions for each permissible transition from a given state $k$. This choice is primarily driven by practical considerations. In many applications, different transitions may depend on different covariate sets or require different functional forms, in which case separate transition-specific models provide a simple and modular implementation. Even when the covariate set is the same across transitions, we still require transition-specific smooth terms (in particular, the baseline $f_{kl}(t)$), and our semi-structured estimation with orthogonalisation is implemented naturally via a binary logistic loss within the available software framework. Specifically, for state $k$, we fit up to $K\!-\!1$ binary models, restricted to observed transitions that are allowed by the process. This yields the permissible binary probabilities $q_{ikl}(t)$ for each borrower $i$.

The link between the multinomial model and separate logit models follows from the equivalence of fitting a logit model based on data from only two states and fitting a logit model conditional on the subject being in one of these two states, as described in \citet[Ch.~7]{agresti2003categorical}. Specifically, it holds that binary and multinomial parameterisations share the same log-odds, that is, 
\begin{equation}
\log \frac{q_{ikl}(t)}{1-q_{ikl}(t)}
= \log \!\left(\frac{\pi_{ikl}(t)/\{\pi_{ikl}(t)+\pi_{ikk}(t)\}}{\pi_{ikk}(t)/\{\pi_{ikl}(t)+\pi_{ikk}(t)\}}\right)
= \log \frac{\pi_{ikl}(t)}{\pi_{ikk}(t)}.
\label{eq:bin}
\end{equation}
However, when a state has multiple permissible exits, the binary probabilities must be transformed to recover competing transition probabilities. We therefore map the predicted $q_{ikl}(t)$ from the binary models to the competing $\pi_{ikl}(t)$; see Section~\ref{subsec:competing}.

A common simplifying assumption is time-homogeneity, i.e., $\pi_{ikl}(t)=\pi_{ikl}$, in which case $\eta_{ikl}$ is time-independent as well, and often one assumes linearity in the covariates:
$\eta_{ikl}=\beta_{kl0}+\sum_{j=1}^p x_{ij}\beta_{klj}$, with intercept $\beta_{kl0}$ and coefficients $\beta_{klj}$ for the time-independent covariates $\bm{x}_i$. However, in our application, a time-nonhomogeneous setting is more suitable. Thus, we include a time-varying baseline term and exogenous covariates $\bm{x}_i(t)$, both potentially transition-specific and time-dependent, as we describe next.

\subsection{Semi-structured multi-state model} \label{subsec:semi-str}
We model the transition-specific predictor $\eta_{ikl}(t)$ as an additive decomposition of covariate effects for subject $i$ at duration $t$. Specifically,  the predictor is decomposed into a structured and an unstructured component. The structured part consists of an intercept $\beta_{kl0}$, linear effects $\sum_{j=1}^p x_{ij}(t)\beta_{klj}$ based on $\bm{x}_i(t)$, and smooth nonlinear effects $\sum_{j'=1}^r f_{klj'}(z_{ij'})$, where $f_{klj'}(\cdot)$ denotes a univariate smooth function of $z_{ij'}\in\mathbb{R}$ (e.g., duration $t$ for baseline terms). The unstructured component captures residual patterns through functions $\sum_{j''=1}^w d_{klj''}(\bm{u}_{ikl})$ of covariates $\bm{u}_{ikl}$ (a subset of $\bm{x}_i(t)$ and/or additional variables). In our implementation, we parameterise each unstructured term as
$d_{klj''}(\bm{u}_{ikl})=\hat{\bm{u}}_{iklj''}^\top \bm{\gamma}_{klj''}$,
where $\hat{\bm{u}}_{iklj''}$ are latent features learned by a deep neural network (DNN) and $\bm{\gamma}_{klj''}$ are the corresponding coefficients. 

This decomposition of the predictor into its structured and unstructured components, as we do, is based on the idea of semi-structured additive predictors proposed by  \citet{rugamer2024semi} in a distributional regression context. Related ideas have since been adapted to survival settings---for example, \citet{kopper2022deeppamm} model complex hazard structures and \citet{medina2024deep} accommodate long-term survivors---and we extend this formulation here to the multi-state case.

In summary, the predictor is specified as:
\begin{equation}
  \eta_{ikl}(t) = 
  \underbrace{\beta_{kl0}+\sum_{j=1}^p x_{ij}(t)\beta_{klj}+\sum_{j'=1}^{r}f_{klj'}(z_{ij'})}_{\eta_{ikl}^{str}\text{ (structured)}} 
  + 
  \underbrace{\sum_{j''=1}^w d_{klj''}(\bm{u}_{ikl})}_{\eta_{ikl}^{unstr}\text{(unstructured)}}.
  \label{eq:eta_full}
\end{equation}
The structured terms $f_{klj'}(\cdot)$, typically represented via basis expansions (e.g., B-splines or cubic regression splines), allow baseline terms to be modelled nonlinearly. Random effects (e.g., subject-specific heterogeneity) can also be added as ridge-penalised linear effects. If available, one can also add, e.g., spatial effects or varying coefficients \citep[see][for details]{wood2017generalized}.

The unstructured terms $d_{klj''}(\bm{u}_{ikl})$ leverage neural network architectures and their efficient training. They can accommodate complex nonlinearities and higher-order interactions among correlated tabular covariates that are cumbersome to capture with purely structured smooth terms, and they also provide a natural route to incorporate unstructured inputs (e.g., text or images) when available.

\subsection{Identifiability and Estimation} \label{subsec:identifiability}
When covariates appear in both the structured term $\eta_{ikl}^{\mathrm{str}}$ and the unstructured term $\eta_{ikl}^{\mathrm{unstr}}$ in \eqref{eq:eta_full}, identifiability of the decomposition is essential. This overlap is common in practice. In many applications, the same covariates are used in both components so that the structured part captures the main, interpretable effects (e.g., approximately linear trends and smooth univariate relationships), while the unstructured component captures remaining departures such as nonlinearities and higher-order interactions among the same inputs. This avoids committing in advance to a specific interaction structure while preserving a transparent structured summary.

Let
\[
\bm{\eta}_{kl}\coloneqq \mathrm{vec}\!\big([\bm{\eta}_{1kl},\dots,\bm{\eta}_{Nkl}]\big)\in\mathbb{R}^{\tilde N},
\qquad
\bm{\eta}_{ikl}=(\eta_{ikl}(0),\dots,\eta_{ikl}(T_i))^\top\in\mathbb{R}^{1+T_i},
\qquad
\tilde N = N + \sum_{i=1}^N T_i .
\]

To ensure identifiability, we construct orthogonal projection matrices $\bm{\mathcal P},\bm{\mathcal P}^\perp\in\mathbb{R}^{\tilde N\times \tilde N}$ such that
$\bm{\eta}_{kl}^{\mathrm{str}}=\bm{\mathcal P}\,\bm{\eta}_{kl}$ and
$\bm{\eta}_{kl}^{\mathrm{unstr}}=\bm{\mathcal P}^\perp \bm{\eta}_{kl}$.
Let $\tilde{\mathbf X}_{\mathrm{str}}\in\mathbb{R}^{\tilde N\times m}$ denote the full design matrix of all structured effects (including the intercept and the bases for spline and random-effect terms, as applicable). We compute a QR factorisation $\tilde{\mathbf X}_{\mathrm{str}}=\mathbf Q\mathbf R$ with
$\mathbf Q\in\mathbb{R}^{\tilde N\times m}$ orthonormal and $\mathbf R\in\mathbb{R}^{m\times m}$ upper triangular.
The orthogonal projector onto the structured column space is
\[
\bm{\mathcal P} \;=\; \mathbf{Q}\mathbf{Q}^\top,
\qquad
\bm{\mathcal P}^\perp \;=\; \mathbf{I}_{\tilde N} - \mathbf{Q}\mathbf{Q}^\top.
\]
This enforces mutual orthogonality of structured and unstructured components and thereby addresses potential identifiability issues \citep{rugamer2024semi}.

For the construction of $\tilde{\mathbf X}_{\mathrm{str}}$, we can write
\[
\tilde{\mathbf X}_{\mathrm{str}}
= \big[\ \mathbbm{1}_{\tilde N}\ \big|\ \mathbf{X}_{\mathrm{lin}}\ \big|\ \mathbf{X}_{\mathrm{spl}}\ \big|\ \mathbf{X}_{\mathrm{re}}\ \big],
\]
where $\mathbbm{1}_{\tilde N}$ is the intercept column, $\mathbf{X}_{\mathrm{lin}}$ stacks the time-indexed linear covariates
$(\bm{x}_1(0)^\top,\dots,\bm{x}_1(T_1)^\top,\dots,\bm{x}_N(T_N)^\top)^\top$,
$\mathbf{X}_{\mathrm{spl}}$ contains bases for structured nonlinear terms (e.g., smoothers of duration $t$ or other covariates), and $\mathbf{X}_{\mathrm{re}}$ holds design columns for random effects when used.

We reformulate estimation as six separate binary logistic regressions to obtain the permissible transitions $q_{ikl}(t)$ (cf.\ Figure~\ref{fig:trans} and  \eqref{eq:bin}). Each regression is implemented with the \texttt{deepregression} \textsf{R} package \citep{JSSv105i02}, using the negative log-likelihood of a logit model as the loss. The approach embeds the entire additive predictor (structured and unstructured terms) within a neural network. The package builds on \texttt{Keras} and \texttt{TensorFlow}, enabling advanced deep-learning features. For efficient matrix handling---particularly the orthogonal decomposition detailed above---mini-batch optimisation performs well in practice \citep{rugamer2024semi}, making estimation reliable within the deep-learning optimisation routines.

\subsection{Competing transition probabilities} \label{subsec:competing}
Our data are observed monthly, with delinquency status updated at month-end. This observation scheme aligns naturally with the discrete-time multi-state model in Section~\ref{subsec:multi-state}. To this end, we transform the probabilities $q_{ikl}(t)$  into competing transition probabilities $\pi_{ikl}(t)$.

Some studies perform this conversion via continuous-time approximations \citep[e.g.,][]{Djeundje2018,djeundje2025devil}. In a discrete-time setting, however, exact transformations can be derived directly from \eqref{eq:bin}, eliminating approximation error. Section~\ref{subsec:competingcom} compares exact and approximate transformations and shows that approximation noise can degrade predictive performance.

For states with a single permissible exit, no adjustment is needed:
\begin{equation}\label{eq:transf_disc1}
\pi_{i01}(t)=q_{i01}(t).
\end{equation}
For states with multiple exits, the transformation rescales the binary probabilities to account for competition. Define the normalising terms
\begin{equation*}
D_{i1}(t)=1-q_{i10}(t)q_{i12}(t),
\end{equation*}
and
\begin{equation*}
D_{i2}(t)=1-q_{i20}(t)q_{i23}(t)-q_{i20}(t)q_{i21}(t)-q_{i21}(t)q_{i23}(t)
+2q_{i20}(t)q_{i21}(t)q_{i23}(t).
\end{equation*}
Then, the exact competing transition probabilities are:
\begin{equation}\label{eq:transf_disc2}
\begin{aligned}
\pi_{i10}(t)&=q_{i10}(t)\frac{1-q_{i12}(t)}{D_{i1}(t)}, &
\pi_{i12}(t)&=q_{i12}(t)\frac{1-q_{i10}(t)}{D_{i1}(t)},\\[4pt]
\pi_{i20}(t)&=q_{i20}(t)\frac{(1-q_{i21}(t))(1-q_{i23}(t))}{D_{i2}(t)}, &
\pi_{i21}(t)&=q_{i21}(t)\frac{(1-q_{i20}(t))(1-q_{i23}(t))}{D_{i2}(t)},\\[4pt]
\pi_{i23}(t)&=q_{i23}(t)\frac{(1-q_{i20}(t))(1-q_{i21}(t))}{D_{i2}(t)}. &&
\end{aligned}
\end{equation}

Given $\pi_{ikl}(t)$, the one-step transition probability matrix for subject $i$ at month $t$ is
\begin{equation*}
  \boldsymbol{P}_i(t)=\left[\begin{array}{cccc}
  \left(1-\pi_{i 01}(t)\right) & \pi_{i 01}(t) & 0 & 0 \\
  \pi_{i 10}(t) & \left(1-\pi_{i 10}(t)-\pi_{i 12}(t)\right) & \pi_{i 12}(t) & 0 \\
  \pi_{i 20}(t) & \pi_{i 21}(t) & \left(1-\pi_{i 20}(t)-\pi_{i 21}(t)-\pi_{i 23}(t)\right) & \pi_{i 23}(t) \\
  0 & 0 & 0 & 1
  \end{array}\right]
\end{equation*}
where state~3 (default) is absorbing.

These one-step matrices can be compounded to obtain the $t_1 \to t_2$ transition matrix
\[
\boldsymbol{P}_i(t_1,t_2)=\prod_{t=t_1+1}^{t_2}\boldsymbol{P}_i(t),
\qquad \boldsymbol{P}_i(t,t)=\mathbf I.
\]
Finally, the distribution over states at time $t_2$ given the observed state $Z_i(t_1)$ at time $t_1$ is
\begin{equation}\label{eq:subj_prob}
\mathbf{p}_{Z_i}(t_1,t_2)=\bigl(\mathbbm{1}\{Z_i(t_1)=0\},\,\mathbbm{1}\{Z_i(t_1)=1\},\,\mathbbm{1}\{Z_i(t_1)=2\},\,\mathbbm{1}\{Z_i(t_1)=3\}\bigr)\,\boldsymbol{P}_i(t_1,t_2).
\end{equation}

\section{Simulation}\label{sec:sim}
This section evaluates the proposed framework in controlled settings. We consider two simulation exercises. The first examines whether the semi-structured model can recover structured and unstructured effects from synthetic data, how recovery improves with sample size, and whether orthogonalisation helps isolate the structured component. The second compares the exact discrete-time transformation of binary transition models to a continuous-time approximation used in related work.

\subsection{Recovering structured and unstructured effects}\label{subsec:sim_recovery}

The simulation design is motivated by the empirical setting in Section~\ref{sec:data}, which comprises four delinquency states and six permissible transitions (Figure~\ref{fig:trans}). We simulate datasets with three sample sizes---10{,}000, 50{,}000, and 100{,}000 loans---each observed monthly over 36 months.

\begin{figure}
	\centering
	\includegraphics[width=0.85\textwidth]{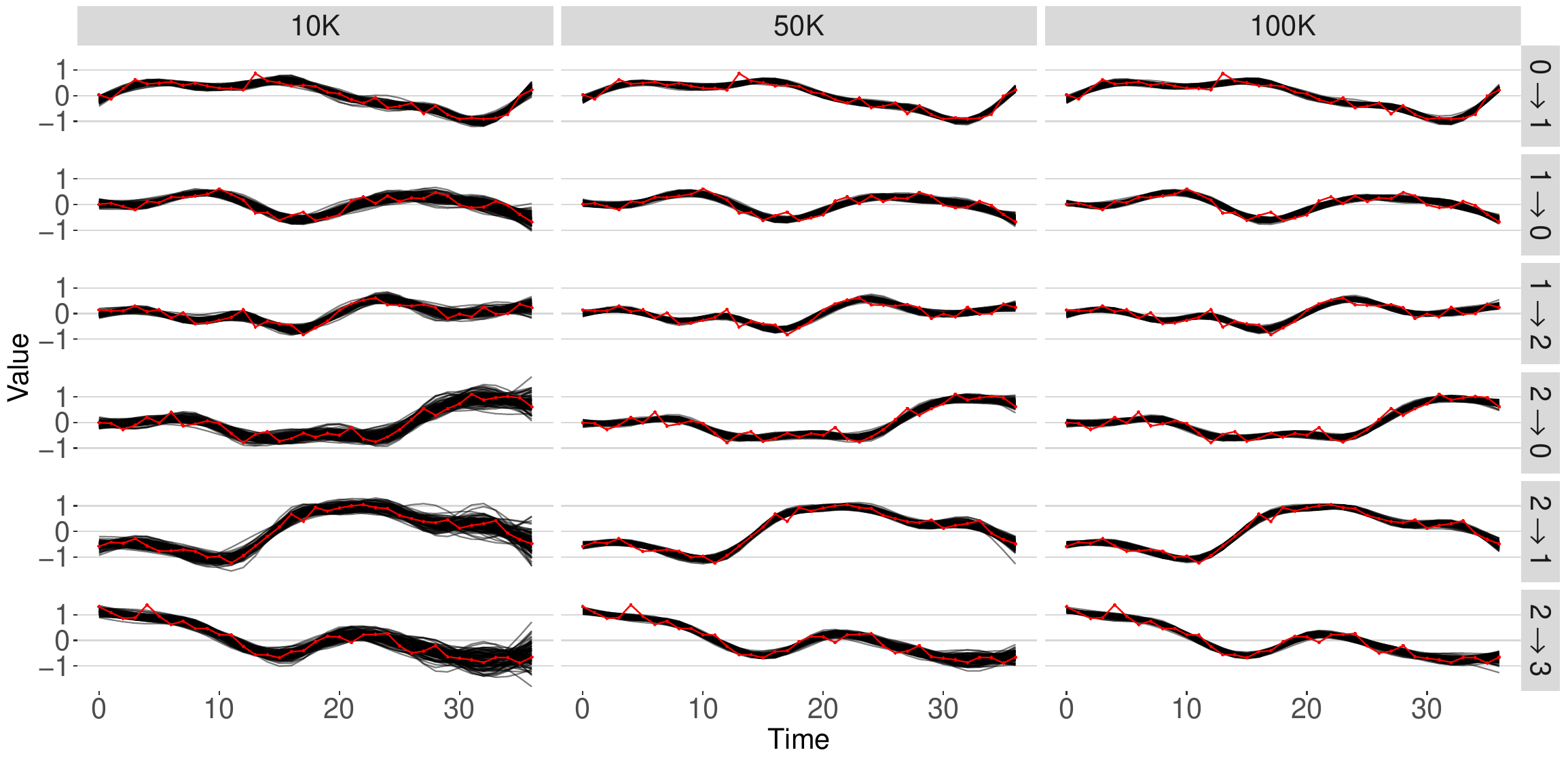}
  \caption{Baseline recovery. Facets are arranged with transitions in rows and sample sizes in columns. Red curves show the true baseline functions; black curves are estimates from 100 simulation runs.}
  \label{fig:baseline}
\end{figure}

\paragraph{Data-generating process.}
For each permissible transition $k \to l$, the baseline term $f_{kl1}(t)$ is generated as a realisation of a second-order random walk. This yields smooth, flexible trajectories, as illustrated by the red curves in Figure~\ref{fig:baseline}.

\begin{figure}
	\centering
	\includegraphics[width=0.85\textwidth]{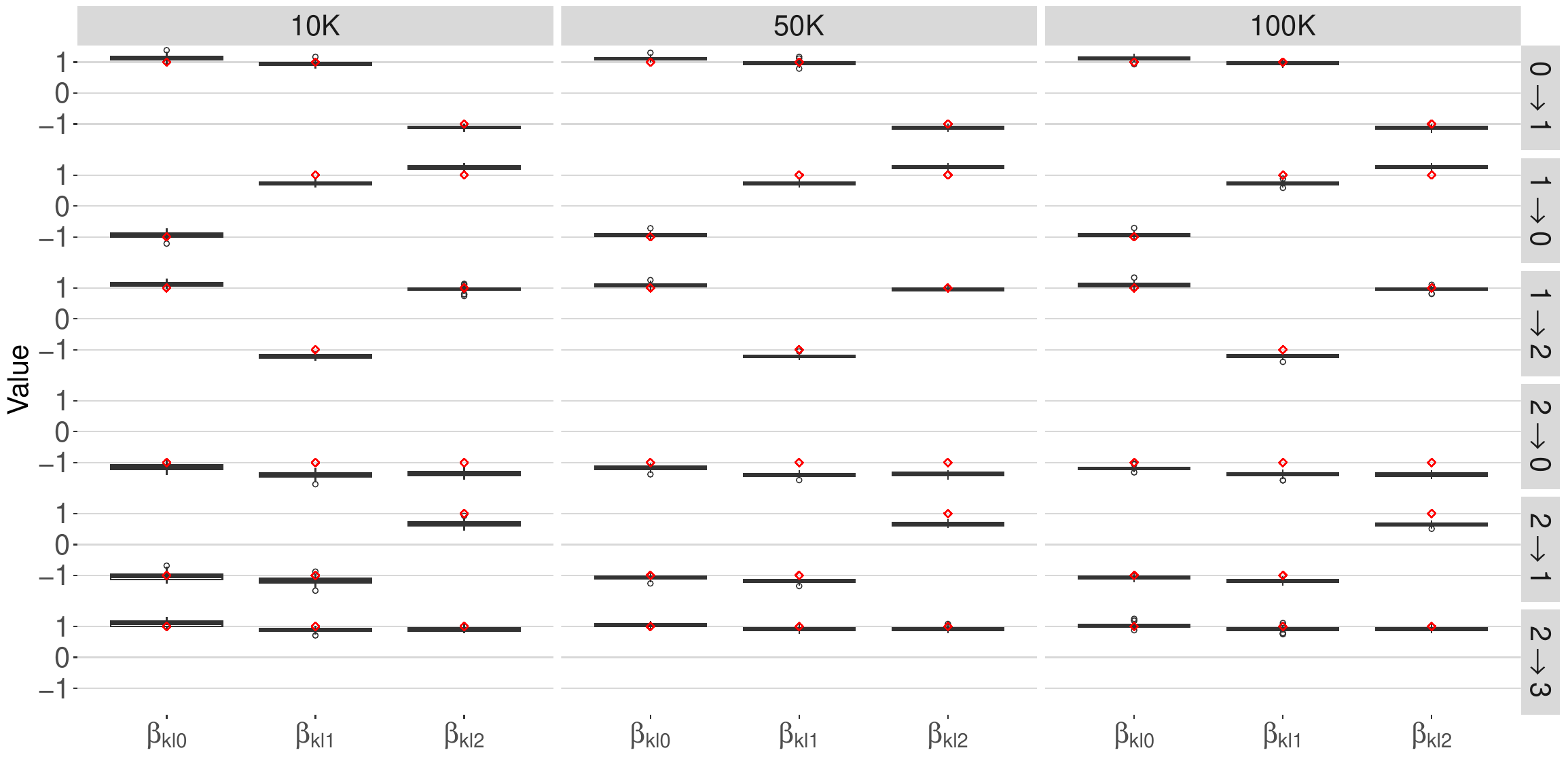}
	\caption{Linear-effect recovery (same facet layout as Figure~\ref{fig:baseline}). Red points indicate the true coefficients; boxplots summarise estimates from 100 simulation runs.}
  \label{fig:lin}
\end{figure}

We include two linear covariate effects, for $x_1$ and $x_2$, together with an intercept. To avoid confounding between the intercept and the baseline term, the baseline curves are generated under a sum-to-zero constraint. With six permissible transitions, this produces 18 linear coefficients in total (three per transition: $\beta_{kl0}$, $\beta_{kl1}$, and $\beta_{kl2}$). The true values are shown as red points in Figure~\ref{fig:lin}.

\begin{figure}
	\centering
	  \includegraphics[width=0.85\textwidth]{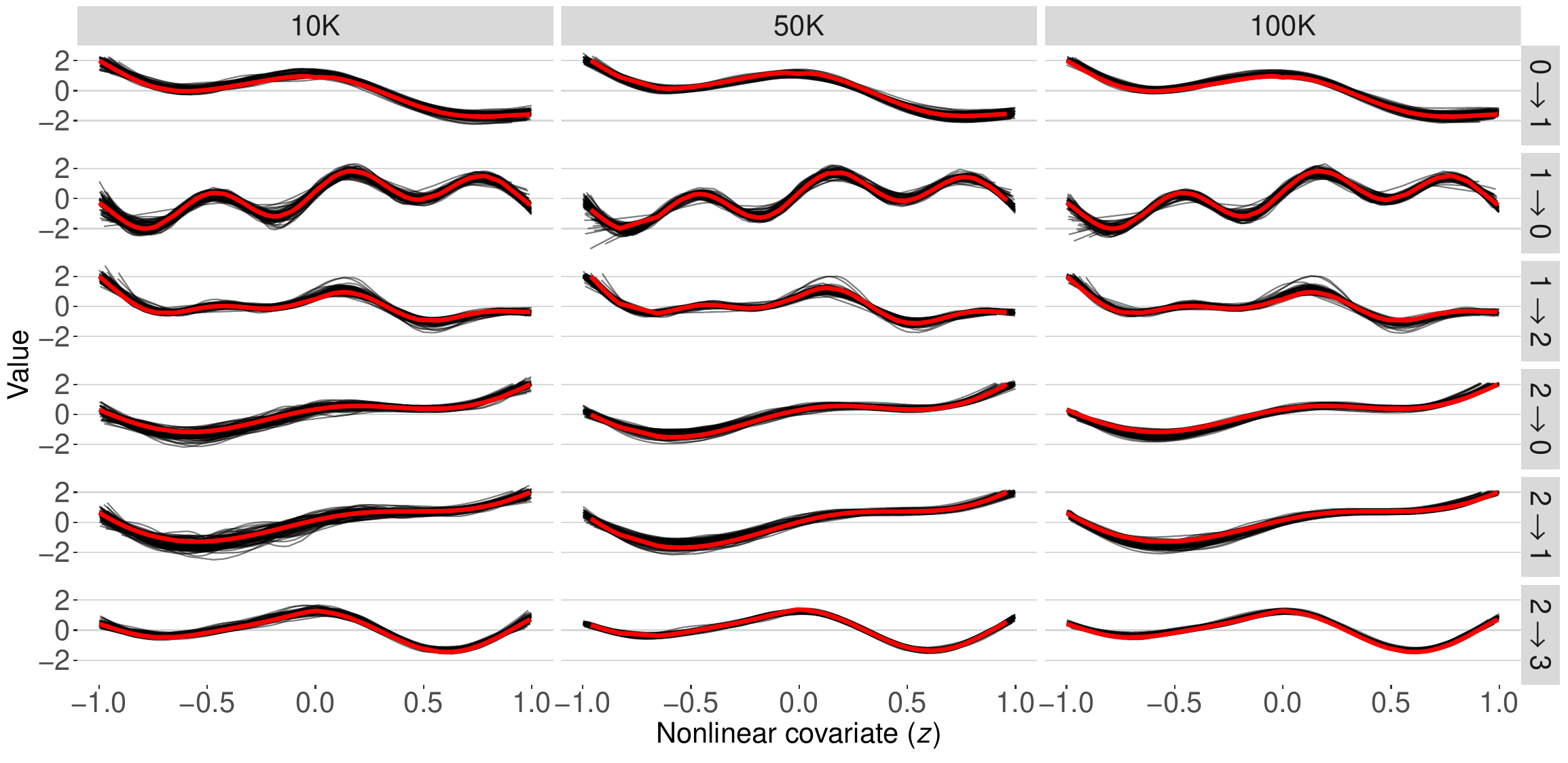}
	  \caption{Nonlinear-effect recovery (same facet layout as Figure~\ref{fig:baseline}). Red curves show the true functions; black curves are estimates from 100 simulation runs.}
  \label{fig:nonlin}
\end{figure}

Nonlinear effects are also included for each transition via an additional covariate $z$, denoted $f_{kl2}(z)$. These effects are drawn from the set of ten nonlinear functions described in Appendix~B of \citet{rugamer2024semi}. The true curves are shown in red in Figure~\ref{fig:nonlin}.

In addition, we include an interaction term $x_1x_2$. All covariates are drawn independently from a uniform distribution on $[-1,1]$. The resulting transition-specific predictor is

\begin{equation*}
  \eta_{kl}(t) \;=\; \beta_{kl0} + \beta_{kl1} x_1 + \beta_{kl2} x_2
  + f_{kl1}(t) + f_{kl2}(z) + x_1 x_2 .
\end{equation*}

\paragraph{Training.}
Baseline and nonlinear effects are estimated using univariate penalised cubic regression splines with basis dimension 10 and evenly spaced knots over the covariate range \citep{wood2017generalized}. The unstructured component is modelled by a neural network with two hidden layers (100 and 50 neurons), ReLU activations \citep{Goodfellow-et-al-2016}, and dropout layers with rate 25\%. This architecture is deliberately chosen to be highly expressive and, in principle, flexible enough to capture most of the data-generating process on its own. It therefore provides a demanding test of whether the orthogonalisation step prevents the neural component from absorbing structured signal and enables recovery of the true structured effects.

Models are trained using stochastic gradient descent (SGD) with a learning rate of 0.01 and a batch size of 32. Early stopping is applied when the validation loss fails to improve for 20 consecutive epochs (patience criterion).\footnote{For the empirical results, we use the same training protocol as a starting point. In local experiments, however, we found that the GeLU activation, $\ell_2$ kernel regularisation, and the Adam optimiser provided slightly better performance.}

\paragraph{Evaluation.}
Figure~\ref{fig:baseline} shows recovery of the baseline terms, while Figures~\ref{fig:lin} and \ref{fig:nonlin} show recovery of the linear and nonlinear effects, respectively. In all plots, true parameter values are highlighted in red. Each model is fitted 100 times using different random seeds. In Figures~\ref{fig:baseline} and \ref{fig:nonlin}, the 100 fitted curves are shown as black lines; for the linear effects, we report boxplots of the estimated coefficients. Across figures, the estimates concentrate around the true values, with concentration tightening as sample size increases.

\subsection{Comparing transformation methods for transition probabilities} \label{subsec:competingcom}
The transformation method used by \citet{Djeundje2018} treats time as continuous and employs an approximation commonly used in Actuarial Mathematics for life contingent risks. Specifically, for the states defined in our delinquency model, the competing transition probabilities between states from time $t-1$ to $t$ are as follows:
\begin{equation}\label{eq:transf_cont}
  \begin{aligned}
      & \pi_{i 01}(t)=q_{i 01}(t) \\
      & \pi_{i 10}(t)=q_{i 10}(t)\left(1-\frac{1}{2} q_{i 12}(t)\right) \\
      & \pi_{i 12}(t)=q_{i 12}(t)\left(1-\frac{1}{2} q_{i 10}(t)\right) \\
      & \pi_{i 20}(t)=q_{i 20}(t)\left(1-\frac{1}{2}\left(q_{i 21}(t)+q_{i 23}(t)\right)+\frac{1}{3} q_{i 21}(t) q_{i 23}(t)\right) \\
      & \pi_{i 21}(t)=q_{i 21}(t)\left(1-\frac{1}{2}\left(q_{i 20}(t)+q_{i 23}(t)\right)+\frac{1}{3} q_{i 20}(t) q_{i 23}(t)\right) \\
      & \pi_{i 23}(t)=q_{i 23}(t)\left(1-\frac{1}{2}\left(q_{i 20}(t)+q_{i 21}(t)\right)+\frac{1}{3} q_{i 20}(t) q_{i 21}(t)\right)
  \end{aligned}
  \end{equation}

However, as noted in Section~\ref{subsec:competing}, observations are recorded monthly and, by construction, state transitions occur between month-ends. The problem is therefore intrinsically discrete. Here, we compare the predictions obtained using the transformation described in \eqref{eq:transf_disc2} with those from \eqref{eq:transf_cont}.

For the simulated data described above (Section~\ref{subsec:sim_recovery}), we randomly select 1000 subjects per setting and calculate their true cumulative transition probabilities starting from state 0 at time 0 to the other three states (1, 2, and 3) ($\mathbf{p}_{Z_i}(0,T)$ in \eqref{eq:subj_prob}) during the length of the study ($T=36$). Table \ref{tab:comp_trans} shows the mean squared error (MSE) and mean absolute error (MAE) of this analysis. We observe that the induced error when using the continuous transformation approach is consistently greater than when the discrete case is used. This reinforces the idea that there is no need to use continuous approximation when the problem is intrinsically discrete.

\begin{table}[h]
    \resizebox{\textwidth}{!}{%
      \begin{tabular}{lcccccc}
        \toprule
        & \multicolumn{3}{c}{MSE (continuous-time)} & \multicolumn{3}{c}{MSE (discrete-time)} \\
        \cmidrule(lr){2-4} \cmidrule(lr){5-7}

  Transition & 10K & 50K & 100K & 10K & 50K & 100K \\ 
    \hline
  0 $\to$ 0 & 0.0016 (0.00024) & 0.0015 (0.00026) & 0.0015 (0.00022) & 0.0007 (0.0002) & 0.0006 (0.00017) & 0.0006 (0.00015) \\ 
  0 $\to$ 1 & 0.0013 (0.00017) & 0.0013 (0.00014) & 0.0012 (0.00014) & 0.0006 (0.00013) & 0.0005 (0.00013) & 0.0005 (0.00012) \\ 
  0 $\to$ 2 & 0.0006 (0.00006) & 0.0006 (0.00005) & 0.0006 (0.00005) & 0.0002 (0.00003) & 0.0001 (0.00003) & 0.0001 (0.00003) \\ 
  0 $\to$ 3 & 0.006 (0.00075) & 0.0057 (0.00059) & 0.0055 (0.0006) & 0.0018 (0.00045) & 0.0015 (0.00043) & 0.0015 (0.00038) \\ 
    \hline

\\
        & \multicolumn{3}{c}{MAE (continuous-time)} & \multicolumn{3}{c}{MAE (discrete-time)} \\
        \cmidrule(lr){2-4} \cmidrule(lr){5-7}
        
  Transition & 10K & 50K & 100K & 10K & 50K & 100K \\ 
    \hline
  0 $\to$ 0 & 0.0215 (0.00174) & 0.0212 (0.00179) & 0.0207 (0.00155) & 0.0129 (0.00171) & 0.0115 (0.00155) & 0.0116 (0.00138) \\ 
  0 $\to$ 1 & 0.021 (0.00138) & 0.0205 (0.00112) & 0.0202 (0.00118) & 0.012 (0.0013) & 0.011 (0.00136) & 0.0109 (0.00133) \\ 
  0 $\to$ 2 & 0.0136 (0.00079) & 0.0134 (0.00063) & 0.0132 (0.00061) & 0.0065 (0.00061) & 0.006 (0.00065) & 0.0059 (0.00061) \\ 
  0 $\to$ 3 & 0.0434 (0.0029) & 0.0424 (0.00234) & 0.0419 (0.00235) & 0.0223 (0.00287) & 0.0201 (0.00296) & 0.0199 (0.00282) \\ 
    \hline
\\
      \end{tabular}%
    } 
    \caption{MSE and MAE comparison for continuous and discrete-time transformation methods.}
    \label{tab:comp_trans}
  \end{table}

\section{Empirical results} \label{sec:emp_res}
We evaluate the model on the Freddie Mac Loan-Level Dataset described in Section~\ref{sec:data}, with training and test sets of $95{,}893$ and $47{,}189$ loans, respectively. Covariates comprise borrower- and loan-level characteristics and macroeconomic variables, aligning with prior work on multi-state credit models \citep{djeundje2025devil,Bocchio2023}. We report results for two covariate sets: (i) borrower/loan variables only and (ii) the same set augmented with macroeconomic variables.

A natural benchmark is a generalised additive model (GAM) that retains the same structured additive predictor but without the neural (unstructured) term $\eta^{\mathrm{unstr}}$ in \eqref{eq:eta_full}. The benchmark is fitted using \texttt{mgcv} \citep{wood2017generalized}.

Consider the transition-specific predictor $\eta_{ikl}(t)$ from state $k$ to $l$ for borrower $i$ at duration time $t$. Let $\bm{x}_i\in\mathbb{R}^p$ denote the borrower- and loan-level characteristics, $f_{kl}(t)$ the time-dependent baseline term, $d_{kl}(\cdot)$ the unstructured component, and $\bm{m}(t+\tau_i-h)\in\mathbb{R}^s$ the macroeconomic variables evaluated at calendar time $t+\tau_i-h$ with $\tau_i$ the loan origination date and $h$ the corresponding time lag. The predictor at duration time $t$ is

\begin{equation} \label{eq:eta_empirical}
  \eta_{ikl}(t)
  = \beta_{kl0}
  + \sum_{j=1}^p x_{ij}\beta_{klj}
  + f_{kl}(t)
  + d_{kl}(\bm{u}_{ikl})
  + \sum_{q=1}^s m_{iq}(t+\tau_i-h)\tilde \beta_{klq}.
\end{equation}
Next, we describe the model architecture and training procedure (Section \ref{subsec:arch}), present the estimated covariate effects (Section \ref{subsec:covariate_effects}), and compare model performance with respect to the corresponding GAM (Section \ref{subsec:performance}).

\subsection{Model architecture and training procedure} \label{subsec:arch}
For each permissible transition (Figure~\ref{fig:trans}), we train a semi-structured model end-to-end, jointly optimising the structured component (linear and smooth terms) and the unstructured neural component. Training uses mini-batch optimisation with the Adam optimiser \citep{kingma2017adam} and a stepwise exponentially decaying learning-rate schedule.\footnote{Let $\alpha_0$ be the initial learning rate, $\gamma$ the decay rate, and $s$ the decay step. After $t$ parameter updates, $\alpha(t)=\alpha_0\,\gamma^{\lfloor t/s\rfloor}$ \citep{chollet2015keras}.}

The unstructured term $d_{kl}(\cdot)$ is implemented as a feed-forward neural network. This is a modelling choice rather than a limitation of the framework, since more complex architectures could be used when justified by the data modality. To keep modelling and tuning tractable, we make two further design choices. First, the neural network takes the full set of covariates available for a given experiment (either borrower/loan variables only or the augmented set including macroeconomic variables), rather than a transition-specific subset. Second, we use the same network architecture for all permissible transitions. Although transition-specific variable selection and architectural search could improve performance, they would also substantially increase the combinatorial complexity of the optimisation problem and are therefore beyond the scope of this work.

The chosen architecture is a multilayer perceptron with two fully connected hidden layers of 128 and 64 units, respectively, each followed by a GeLU activation \citep{hendrycks2016gaussian}. We apply $\ell_2$ kernel regularisation with penalty $10^{-3}$ to all dense layers. As described in Section~\ref{subsec:semi-str}, the final layer is a single linear unit that contributes additively to the predictor. This architecture was selected after non-exhaustive experimentation with other flexible alternatives and was found to perform robustly across transitions without requiring further transition-specific tuning.

We tune optimisation hyperparameters over the following discrete search space:
\[
\begin{aligned}
\text{initial learning rate } &\in \{0.01,\ 0.00464159,\ 0.00215443,\ 0.001,\ 0.00046416,\ 0.00021544,\ 0.0001\},\\
\text{decay steps } &\in \{5000,\ 10000,\ 15000\},\\
\text{decay rate } &\in \{0.8,\ 0.9,\ 0.95,\ 0.99\},\\
\text{first-moment decay } &\in \{0.8,\ 0.9,\ 0.99\},\\
\text{batch size } &\in \{32,\ 128\}.
\end{aligned}
\]

The initial learning rate is selected from a seven-point logarithmically spaced grid between $10^{-2}$ and $10^{-4}$, given by
$10^{-2}\left(\frac{10^{-4}}{10^{-2}}\right)^{\frac{r-1}{6}}$, $r=1,\dots,7$. Furthermore, in Adam, the first-moment decay parameter controls the exponential moving average of past gradients, and larger values yield smoother (more persistent) momentum estimates \citep{kingma2017adam}.

From this grid, we sample 200 configurations uniformly at random (except for the $0\!\to\!1$ transition; see below). For each configuration, we train two independent initialisations (different random seeds) and apply early stopping with a patience of 10 epochs. We select the configuration that minimises the mean validation loss across the two runs, and then refit the selected configuration on the full training data using the same learning-rate schedule.

Due to computational constraints, hyperparameter tuning for the $0\!\to\!1$ transition is conducted using 10 random draws from the same search space. However, the remainder of the protocol is unchanged. The hyperparameters that minimise validation loss across these trials are then used to retrain on the full training set.

The final hyperparameter configurations per transition are summarised in Table~\ref{tab:hyperparams}.

\begin{table}[ht]
  \centering
  \caption{Final hyperparameter configurations for each transition model.}
  \label{tab:hyperparams}
  \resizebox{\textwidth}{!}{%
  \begin{tabular}{lccccccccc}
      \toprule
      \textbf{\#} & \textbf{Macro. Vars.} & \textbf{Transition} & \textbf{Initial LR} & \textbf{Decay Steps} & \textbf{Decay Rate} & \textbf{Optimizer} & \textbf{First-moment} & \textbf{Epochs} & \textbf{Batch Size} \\ 
      \midrule
       1  & No  & 0 $\to$ 1 & 0.00046416 & 10,000 & 0.99 & Adam & 0.80 & 37 & 32 \\
       2  & No  & 1 $\to$ 0 &    0.01 & 10,000 & 0.80 & Adam & 0.99 & 40 & 32 \\
       3  & No  & 1 $\to$ 2 &    0.01 & 15,000 & 0.99 & Adam & 0.80 & 27 & 32 \\
       4  & No  & 2 $\to$ 0 &    0.01 & 15,000 & 0.90 & Adam & 0.99 & 47 & 32 \\
       5  & No  & 2 $\to$ 1 &    0.01 & 10,000 & 0.80 & Adam & 0.99 & 99 & 32 \\
       6  & No  & 2 $\to$ 3 &    0.01 & 15,000 & 0.99 & Adam & 0.90 & 39 & 32 \\
       7  & Yes  & 0 $\to$ 1 & 0.00464159 & 10,000 & 0.95 & Adam & 0.80 & 9 & 32 \\
       8  & Yes  & 1 $\to$ 0 &    0.01 & 10,000 & 0.95 & Adam & 0.99 & 29 & 32 \\
       9  & Yes  & 1 $\to$ 2 &    0.01 & 10,000 & 0.80 & Adam & 0.99 & 19 & 32 \\
      10  & Yes  & 2 $\to$ 0 &    0.01 & 15,000 & 0.99 & Adam & 0.80 & 35 & 32 \\
      11  & Yes  & 2 $\to$ 1 &    0.01 & 5,000 & 0.80 & Adam & 0.90 & 122 & 32 \\
      12  & Yes  & 2 $\to$ 3 &    0.01 & 10,000 & 0.95 & Adam & 0.99 & 55 & 32 \\
      \bottomrule
  \end{tabular}%
  }
\end{table}

\subsection{Covariate effects}\label{subsec:covariate_effects}
We summarise estimated \emph{linear} effects of borrower- and loan-level covariates across the six permissible transitions, represented by the coefficients $\beta_{klj}$, $j=0,\dots,p$, in \eqref{eq:eta_empirical}. Results are reported for specifications without and with macroeconomic variables (the $\tilde{\beta}_{klq}$, $q=1,\dots,s$, in \eqref{eq:eta_empirical}), and for both the GAM and the semi-structured model. Each cell reports the mean and a 95\% interval. For the semi-structured models, intervals are based on a 100-resample nonparametric bootstrap \citep{efron1994introduction}. For reference, Table~\ref{tab:covariate_dictionary} in Appendix~\ref{app:dict} maps coefficient names to the underlying variables and category definitions.

\subsubsection{Models without macroeconomic variables}
\paragraph{Linear effects.}
Tables~\ref{tab:lin_eff_gam_mac0} and~\ref{tab:lin_eff_deep_mac0} report linear effects for the GAM and the semi-structured model, respectively. Effects are on the log-odds scale for each transition (columns).

Overall, both models agree on the main directional patterns. Entry into delinquency ($0\!\to\!1$) is associated with lower credit quality (negative effects for \texttt{fico}) and higher leverage/affordability risk (positive effects for \texttt{ltv} and higher DTI categories). The occupancy indicators \texttt{occpy\_stsP} (principal residence) and \texttt{occpy\_stsS} (second home) are negative for cure transitions ($1\!\to\!0$ and $2\!\to\!0$) and also negative for deterioration/default transitions ($1\!\to\!2$ and $2\!\to\!3$). Relative to investment properties, this pattern suggests greater state persistence, i.e.\ conditional on being delinquent, principal and second-home loans are less likely to cure but also less likely to deteriorate further. The COVID indicator \texttt{cov\_dummy} shows large positive effects for deterioration transitions ($1\!\to\!2$ and $2\!\to\!3$) in both models, consistent with a sharp upward shift in conditional risk during the pandemic period.

Differences between the GAM and semi-structured estimates are most visible for covariates where nonlinearities and interactions are plausible \citep{medina2024deep}. For example, the semi-structured model assigns a larger positive linear effect to \texttt{int\_rt} for delinquency entry ($0\!\to\!1$) and more negative effects for cures ($1\!\to\!0$ and $2\!\to\!0$). By contrast, several DTI effects are attenuated and, for some transitions, shift towards more negative values under the semi-structured specification, consistent with the neural component absorbing part of the complex affordability signal. Effects for \texttt{ltv} are also generally smaller in magnitude under the semi-structured model, particularly for $1\!\to\!2$ and $2\!\to\!3$, where the GAM produces comparatively large coefficients with wide uncertainty.

High-cardinality categorical information encoded via WOE-style terms (\texttt{seller\_name\_te}, \texttt{servicer\_name\_te}, \texttt{us\_state\_te}) is influential in both approaches. Intervals are generally tighter under the semi-structured specification. In contrast, a small number of rare factor levels (e.g., some property-type indicators) produce large estimates with wide intervals for $0\!\to\!1$, likely reflecting a combination of sparse representation in the specific subsample and the reduced hyperparameter search for $0\!\to\!1$.

\begin{landscape}
  \begin{table}[p]
  \centering
  \caption{Linear covariate effects (mean and 95\% interval). GAM without macroeconomic variables.}
  \label{tab:lin_eff_gam_mac0}
  \begin{tabular}{lcccccc}
    \toprule
    \textbf{Name} & 0 $\to$ 1 & 1 $\to$ 0 & 1 $\to$ 2 & 2 $\to$ 0 & 2 $\to$ 1 & 2 $\to$ 3 \\
    \midrule
    channelC & -0.12 (-0.17, -0.07) & 0.06 (-0.04, 0.16) & 0.08 (-0.05, 0.21) & -0.25 (-0.53, 0.02) & 0.22 (-0.14, 0.57) & -0.30 (-0.52, -0.08) \\
    channelR & -0.27 (-0.32, -0.22) & 0.09 (-0.00, 0.19) & 0.24 (0.12, 0.36) & -0.13 (-0.38, 0.12) & 0.08 (-0.26, 0.42) & -0.21 (-0.41, -0.01) \\
    cnt\_borrone & 0.48 (0.45, 0.51) & -0.19 (-0.25, -0.13) & 0.05 (-0.03, 0.13) & -0.02 (-0.18, 0.14) & 0.04 (-0.17, 0.25) & -0.03 (-0.15, 0.10) \\
    cnt\_units & -0.04 (-0.10, 0.02) & 0.17 (0.02, 0.32) & 0.17 (-0.01, 0.34) & 0.01 (-0.36, 0.37) & -0.13 (-0.63, 0.37) & 0.03 (-0.24, 0.29) \\
    cov\_dummy & 1.04 (1.00, 1.08) & -0.15 (-0.23, -0.07) & 1.51 (1.42, 1.61) & 0.14 (-0.05, 0.32) & -0.44 (-0.67, -0.22) & 1.15 (1.00, 1.30) \\
    dti1 & 0.15 (0.07, 0.23) & -0.16 (-0.33, 0.01) & -0.05 (-0.28, 0.17) & -0.31 (-0.80, 0.18) & -0.26 (-0.86, 0.35) & -0.16 (-0.57, 0.25) \\
    dti2 & 0.34 (0.26, 0.41) & -0.20 (-0.36, -0.04) & 0.03 (-0.18, 0.24) & -0.35 (-0.81, 0.12) & -0.27 (-0.85, 0.30) & -0.21 (-0.61, 0.18) \\
    dti3 & 0.48 (0.41, 0.56) & -0.39 (-0.54, -0.23) & -0.03 (-0.24, 0.18) & -0.27 (-0.73, 0.19) & -0.37 (-0.93, 0.20) & -0.18 (-0.57, 0.21) \\
    dti5 & 0.42 (0.32, 0.51) & -0.31 (-0.50, -0.12) & -0.11 (-0.36, 0.15) & -0.47 (-1.03, 0.08) & -0.93 (-1.64, -0.23) & -0.09 (-0.55, 0.36) \\
    fico & -5.73 (-5.87, -5.60) & 2.36 (2.13, 2.59) & 1.43 (1.14, 1.73) & 1.60 (0.98, 2.23) & -0.87 (-1.70, -0.04) & 1.55 (1.06, 2.04) \\
    flag\_fthbY & -0.09 (-0.13, -0.05) & -0.03 (-0.11, 0.06) & 0.12 (0.01, 0.22) & 0.12 (-0.10, 0.33) & 0.13 (-0.14, 0.39) & 0.05 (-0.12, 0.21) \\
    flag\_scY & 0.04 (-0.05, 0.13) & 0.10 (-0.08, 0.28) & -0.07 (-0.29, 0.15) & -0.22 (-0.71, 0.27) & -0.57 (-1.26, 0.13) & -0.34 (-0.69, 0.01) \\
    int\_rt & 1.15 (1.00, 1.30) & -0.39 (-0.69, -0.08) & -0.11 (-0.49, 0.27) & -0.72 (-1.38, -0.05) & 0.05 (-0.78, 0.89) & -0.43 (-0.96, 0.10) \\
    loan\_purposeN & -0.07 (-0.12, -0.02) & 0.03 (-0.06, 0.13) & -0.05 (-0.16, 0.07) & 0.10 (-0.15, 0.36) & 0.23 (-0.10, 0.55) & -0.16 (-0.36, 0.04) \\
    loan\_purposeP & -0.03 (-0.08, 0.01) & 0.12 (0.03, 0.22) & -0.12 (-0.24, -0.00) & 0.13 (-0.13, 0.38) & 0.13 (-0.21, 0.46) & -0.23 (-0.43, -0.04) \\
    ltv & 0.64 (0.23, 1.04) & -0.87 (-1.67, -0.07) & 1.63 (0.61, 2.65) & -1.61 (-3.90, 0.68) & -1.96 (-5.01, 1.08) & 0.35 (-1.39, 2.08) \\
    mi\_cxl\_indN & 0.28 (0.16, 0.40) & -0.37 (-0.61, -0.13) & 0.01 (-0.28, 0.30) & 0.49 (-0.14, 1.12) & 0.94 (0.21, 1.67) & 0.50 (0.01, 0.99) \\
    mi\_cxl\_indY & -0.01 (-0.12, 0.10) & -0.19 (-0.39, 0.02) & -0.08 (-0.33, 0.17) & 0.14 (-0.43, 0.71) & 0.75 (0.09, 1.40) & -0.04 (-0.49, 0.40) \\
    mi\_pct & -0.03 (-0.21, 0.14) & 0.04 (-0.27, 0.34) & -0.22 (-0.59, 0.15) & -0.78 (-1.59, 0.03) & -0.96 (-1.91, -0.02) & -0.61 (-1.24, 0.01) \\
    occpy\_stsP & -0.05 (-0.11, 0.01) & -0.20 (-0.33, -0.07) & -0.22 (-0.37, -0.06) & -0.73 (-1.08, -0.39) & -0.57 (-1.05, -0.09) & -0.60 (-0.90, -0.31) \\
    occpy\_stsS & -0.16 (-0.26, -0.05) & -0.31 (-0.53, -0.08) & -0.37 (-0.65, -0.09) & -0.74 (-1.36, -0.11) & -0.38 (-1.22, 0.47) & -0.47 (-0.97, 0.03) \\
    orig\_loan\_term & -0.02 (-0.10, 0.05) & -0.17 (-0.32, -0.03) & 0.02 (-0.16, 0.20) & 0.16 (-0.21, 0.53) & -0.22 (-0.69, 0.25) & 0.21 (-0.09, 0.51) \\
    orig\_upb & 0.81 (0.66, 0.96) & -0.23 (-0.52, 0.06) & 0.48 (0.12, 0.83) & -0.15 (-0.89, 0.58) & 0.71 (-0.21, 1.62) & 0.49 (-0.07, 1.05) \\
    prop\_typeCP & -0.38 (-0.88, 0.11) & -0.06 (-1.27, 1.15) & 1.22 (-0.04, 2.49) & -0.73 (-2.15, 0.68) & -0.31 (-1.97, 1.34) & -1.31 (-2.72, 0.10) \\
    prop\_typeMH & 0.20 (-0.02, 0.41) & 0.49 (0.01, 0.96) & 0.28 (-0.32, 0.88) & -0.26 (-1.42, 0.90) & -0.31 (-2.10, 1.48) & -0.48 (-1.42, 0.46) \\
    prop\_typePU & 0.10 (0.04, 0.16) & -0.04 (-0.17, 0.09) & -0.07 (-0.23, 0.08) & -0.08 (-0.41, 0.25) & -0.13 (-0.58, 0.33) & -0.14 (-0.40, 0.12) \\
    prop\_typeSF & 0.17 (0.11, 0.22) & -0.13 (-0.25, -0.01) & -0.22 (-0.36, -0.08) & -0.11 (-0.42, 0.20) & -0.10 (-0.52, 0.33) & -0.34 (-0.58, -0.10) \\
    seller\_name\_te & 2.47 (0.14, 4.80) & 2.32 (1.69, 2.95) & 2.84 (2.19, 3.49) & 2.22 (1.04, 3.41) & 4.93 (3.10, 6.77) & 3.46 (2.16, 4.76) \\
    servicer\_name\_te & 5.35 (3.03, 7.66) & 1.62 (0.95, 2.29) & 0.81 (0.06, 1.56) & 3.38 (2.46, 4.29) & 1.40 (-1.13, 3.93) & 2.04 (1.11, 2.98) \\
    us\_state\_te & 3.49 (1.16, 5.82) & 3.48 (2.93, 4.02) & 3.45 (2.94, 3.97) & 4.26 (3.48, 5.04) & 4.93 (3.68, 6.18) & 4.35 (3.41, 5.29) \\
    \bottomrule
  \end{tabular}%
\end{table}

\end{landscape}

\begin{landscape}
  \begin{table}[p]
  \centering
  \caption{Linear covariate effects (mean and 95\% interval). Semi-structured without macroeconomic variables.}
  \label{tab:lin_eff_deep_mac0}
  \begin{tabular}{lcccccc}
    \toprule
    \textbf{Name} & 0 $\to$ 1 & 1 $\to$ 0 & 1 $\to$ 2 & 2 $\to$ 0 & 2 $\to$ 1 & 2 $\to$ 3 \\
    \midrule
    channelC & -0.14 (-0.20, -0.11) & -0.01 (-0.13, 0.10) & 0.08 (-0.07, 0.20) & -0.24 (-0.48, -0.03) & 0.20 (-0.05, 0.50) & -0.16 (-0.37, 0.04) \\
    channelR & -0.27 (-0.32, -0.22) & 0.08 (-0.03, 0.19) & 0.23 (0.10, 0.36) & -0.17 (-0.40, 0.08) & 0.11 (-0.20, 0.42) & -0.21 (-0.41, -0.04) \\
    cnt\_borrone & 0.47 (0.43, 0.51) & -0.20 (-0.26, -0.13) & 0.03 (-0.04, 0.13) & -0.02 (-0.17, 0.13) & 0.06 (-0.17, 0.25) & 0.00 (-0.10, 0.12) \\
    cnt\_units & -0.04 (-0.10, 0.00) & 0.11 (-0.01, 0.26) & 0.09 (-0.11, 0.28) & -0.16 (-0.49, 0.13) & -0.17 (-0.75, 0.25) & -0.10 (-0.28, 0.13) \\
    cov\_dummy & 1.02 (0.98, 1.06) & -0.16 (-0.24, -0.06) & 1.49 (1.39, 1.57) & 0.16 (-0.04, 0.34) & -0.43 (-0.66, -0.22) & 1.20 (1.05, 1.36) \\
    dti1 & -0.04 (-0.10, 0.03) & -0.25 (-0.41, -0.07) & -0.08 (-0.35, 0.18) & -0.39 (-0.77, -0.01) & -0.23 (-0.78, 0.35) & -0.37 (-0.77, -0.03) \\
    dti2 & 0.16 (0.11, 0.22) & -0.25 (-0.43, -0.06) & -0.02 (-0.28, 0.19) & -0.43 (-0.79, -0.09) & -0.30 (-0.76, 0.25) & -0.43 (-0.80, -0.12) \\
    dti3 & 0.31 (0.25, 0.37) & -0.44 (-0.62, -0.25) & -0.17 (-0.38, 0.05) & -0.37 (-0.72, -0.02) & -0.37 (-0.89, 0.15) & -0.27 (-0.65, 0.03) \\
    dti5 & 0.24 (0.16, 0.30) & -0.37 (-0.57, -0.19) & -0.29 (-0.56, -0.05) & -0.59 (-0.99, -0.09) & -0.92 (-1.49, -0.38) & -0.32 (-0.78, 0.09) \\
    fico & -5.33 (-5.41, -5.26) & 2.26 (2.07, 2.49) & 1.27 (1.00, 1.53) & 1.30 (0.75, 1.92) & -0.92 (-1.64, -0.24) & 1.45 (0.97, 1.83) \\
    flag\_fthbY & -0.11 (-0.16, -0.06) & 0.04 (-0.06, 0.12) & 0.17 (0.04, 0.30) & 0.10 (-0.12, 0.32) & 0.14 (-0.11, 0.43) & 0.08 (-0.09, 0.24) \\
    flag\_scY & -0.10 (-0.20, 0.00) & 0.08 (-0.08, 0.25) & -0.10 (-0.32, 0.13) & -0.19 (-0.59, 0.12) & -0.58 (-1.38, 0.20) & -0.34 (-0.69, -0.02) \\
    int\_rt & 1.52 (1.42, 1.60) & -0.55 (-0.81, -0.26) & -0.26 (-0.58, 0.06) & -0.86 (-1.46, -0.29) & -0.02 (-0.90, 0.78) & -0.57 (-1.06, -0.04) \\
    loan\_purposeN & -0.05 (-0.10, 0.00) & -0.01 (-0.13, 0.09) & -0.10 (-0.26, 0.02) & 0.06 (-0.20, 0.33) & 0.17 (-0.12, 0.42) & -0.20 (-0.39, 0.01) \\
    loan\_purposeP & -0.01 (-0.07, 0.04) & 0.14 (0.04, 0.24) & -0.14 (-0.25, -0.03) & 0.10 (-0.12, 0.35) & 0.12 (-0.13, 0.52) & -0.28 (-0.46, -0.09) \\
    ltv & 0.35 (0.19, 0.52) & -1.14 (-1.85, -0.39) & 0.79 (-0.09, 1.75) & -1.25 (-2.72, 0.09) & -1.33 (-3.06, 0.54) & -0.27 (-1.66, 0.92) \\
    mi\_cxl\_indN & 0.22 (0.15, 0.29) & -0.34 (-0.57, -0.05) & 0.06 (-0.18, 0.35) & 0.46 (0.00, 0.91) & 0.94 (0.33, 1.58) & 0.58 (0.16, 1.05) \\
    mi\_cxl\_indY & -0.09 (-0.18, 0.02) & -0.22 (-0.45, 0.00) & -0.07 (-0.26, 0.19) & 0.12 (-0.35, 0.61) & 0.69 (0.13, 1.44) & -0.03 (-0.38, 0.36) \\
    mi\_pct & 0.04 (-0.07, 0.16) & 0.08 (-0.27, 0.41) & -0.18 (-0.55, 0.05) & -0.81 (-1.38, -0.12) & -0.99 (-1.71, -0.28) & -0.70 (-1.25, -0.17) \\
    occpy\_stsP & 0.02 (-0.03, 0.06) & -0.29 (-0.41, -0.14) & -0.29 (-0.46, -0.12) & -0.88 (-1.22, -0.56) & -0.59 (-1.01, -0.16) & -0.77 (-1.12, -0.50) \\
    occpy\_stsS & -0.27 (-0.38, -0.14) & -0.36 (-0.59, -0.15) & -0.46 (-0.79, -0.15) & -0.88 (-1.48, -0.27) & -0.39 (-1.31, 0.51) & -0.65 (-1.17, -0.14) \\
    orig\_loan\_term & -0.09 (-0.17, -0.03) & -0.16 (-0.29, -0.03) & 0.03 (-0.13, 0.22) & 0.09 (-0.25, 0.44) & -0.26 (-0.66, 0.26) & 0.20 (-0.07, 0.47) \\
    orig\_upb & 0.90 (0.76, 1.04) & -0.24 (-0.49, 0.02) & 0.48 (0.05, 0.81) & -0.12 (-0.78, 0.59) & 0.67 (-0.33, 1.58) & 0.52 (0.07, 1.06) \\
    prop\_typeCP & -6.83 (-7.67, -5.95) & 0.01 (-1.43, 1.31) & 1.21 (-0.19, 2.73) & -0.81 (-2.84, 0.29) & -0.39 (-6.01, 1.73) & -1.48 (-5.78, -0.38) \\
    prop\_typeMH & -0.91 (-1.50, -0.49) & 0.47 (-0.01, 1.02) & 0.25 (-0.26, 0.82) & -0.25 (-2.02, 1.48) & -0.27 (-5.83, 1.88) & -0.44 (-1.74, 1.03) \\
    prop\_typePU & 0.02 (-0.04, 0.08) & -0.06 (-0.19, 0.09) & -0.17 (-0.35, -0.01) & -0.16 (-0.54, 0.14) & -0.08 (-0.46, 0.30) & -0.27 (-0.49, -0.02) \\
    prop\_typeSF & 0.09 (0.05, 0.14) & -0.20 (-0.35, -0.10) & -0.24 (-0.38, -0.09) & -0.23 (-0.54, 0.07) & -0.11 (-0.51, 0.23) & -0.40 (-0.62, -0.13) \\
    seller\_name\_te & 3.46 (3.26, 3.66) & 2.19 (1.64, 2.82) & 2.57 (1.98, 3.16) & 2.17 (1.28, 3.19) & 4.58 (2.92, 6.00) & 1.99 (0.92, 2.91) \\
    servicer\_name\_te & 5.22 (4.91, 5.51) & 1.12 (0.48, 1.76) & 0.45 (-0.26, 1.11) & 2.70 (1.91, 3.37) & 1.24 (-0.72, 2.81) & 1.12 (0.32, 1.97) \\
    us\_state\_te & 3.99 (3.74, 4.23) & 2.77 (2.27, 3.29) & 3.03 (2.54, 3.55) & 3.73 (3.17, 4.41) & 4.69 (3.64, 5.70) & 2.34 (1.44, 3.12) \\
    \bottomrule
  \end{tabular}%
\end{table}

\end{landscape}

\begin{figure}[htbp]
	\centering
	  \includegraphics[width=0.85\textwidth]{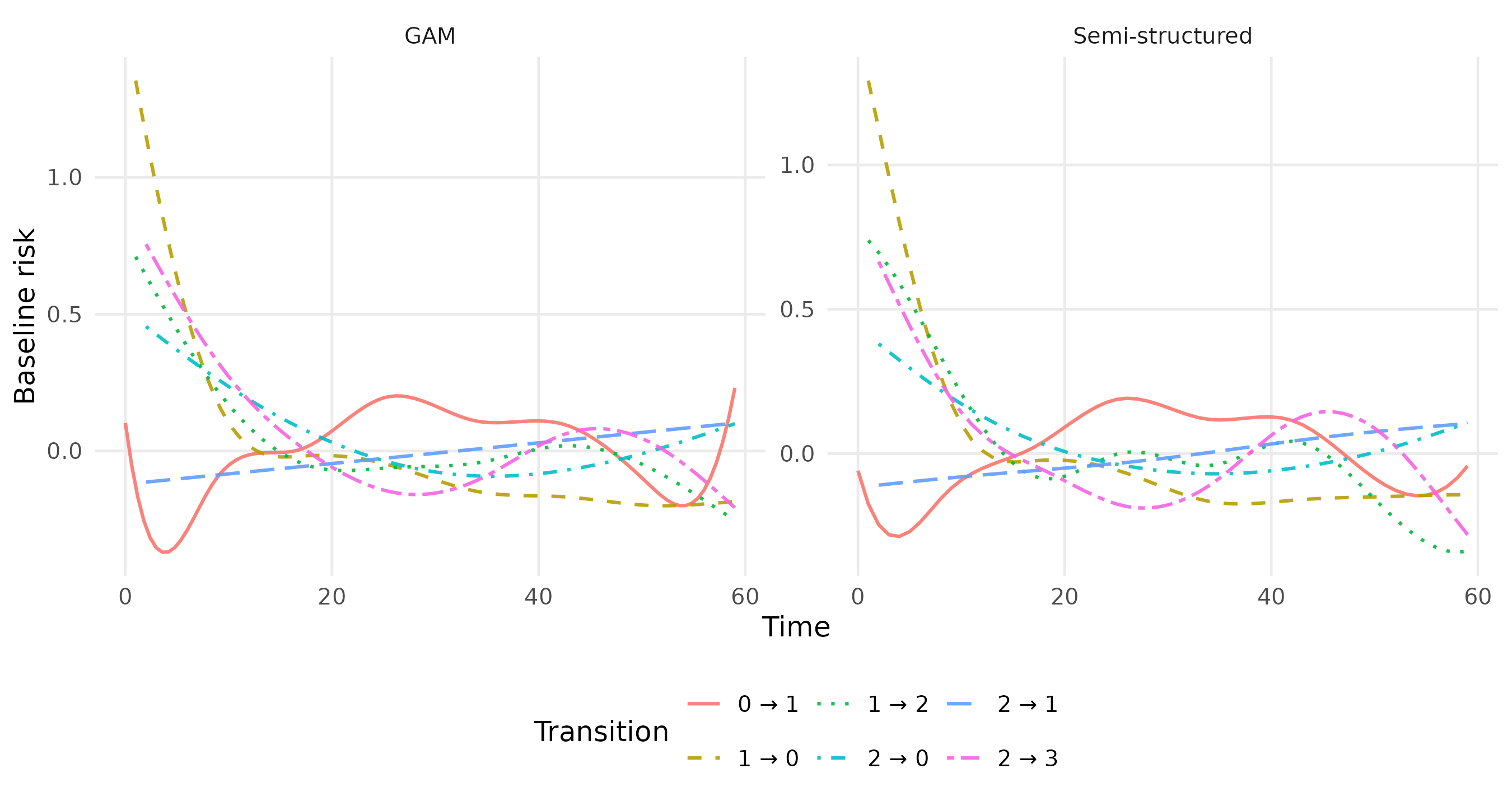}
	  \caption{Estimated time-dependent baseline risk terms $f_{kl}(t)$ (log-odds scale) by transition for the GAM and the semi-structured model, estimated without macroeconomic covariates. Positive values indicate higher baseline propensity for the corresponding transition at time $t$.}
  \label{fig:baseline_mac0}
\end{figure}

\paragraph{Baseline risk terms.}
Figure~\ref{fig:baseline_mac0} summarises the estimated time-dependent baseline terms $f_{kl}(t)$ for the no-macro specification. These smooths capture systematic duration dependence that remains after conditioning on the observed covariates.

Both models exhibit pronounced \emph{duration dependence}, meaning that, after conditioning on borrower- and loan-level covariates, transition probabilities still vary systematically with time since origination, as has also been shown for other portfolios \citep{Djeundje2018,Bocchio2023}. Baseline propensities for cure ($1\!\to\!0$) and deterioration ($1\!\to\!2$) are highest early in the loan life and decline sharply after roughly the first 20 months. Likewise, the transition from $2\!\to\!3$ is most likely shortly after a loan enters state~2 and decreases thereafter, indicating that defaults from severe delinquency tend to occur relatively quickly once that state is reached. In contrast, the baseline propensity for partial improvement ($2\!\to\!1$) increases gradually with duration, suggesting that among loans that remain in state~2, recovery becomes relatively more likely at longer horizons. Estimates in the far tail should be interpreted cautiously because fewer observations support the smooth terms. Overall, the GAM and semi-structured baseline terms are qualitatively similar, implying that the semi-structured model's gains (see Section~\ref{subsec:performance}) arise mainly from how it captures remaining nonlinear and interaction structure through the neural component rather than from materially different duration effects.

\subsubsection{Models with macroeconomic variables} \label{subsubsec:mac1}
\paragraph{Linear effects.}
Tables~\ref{tab:lin_eff_gam_mac1} and~\ref{tab:lin_eff_deep_mac1} in the Appendix~\ref{app:lin_mac1} report results analogous to those above after augmenting the covariate set with macroeconomic indicators. This specification investigates whether aggregate economic conditions add information beyond origination and performance history, and how any additional signal is shared between the structured and unstructured components.

Overall, macro variables show some associations with transition risks, but patterns differ across models. In the GAM case, \texttt{unrt\_lag} and \texttt{cci\_lag} are negative in sign for deterioration/default transitions (notably $0\!\to\!1$, $1\!\to\!2$, and $2\!\to\!3$). In the semi-structured model, \texttt{unrt\_lag} remains negative but is less transition-specific, while \texttt{cci\_lag} is mostly negative with wider intervals and smaller magnitudes, which could indicate that part of the signal is captured nonlinearly by the unstructured component. The largest contrast is seen by \texttt{ipi\_lag} and \texttt{mort30\_lag}, where the GAM produces large, transition-varying effects (especially for \texttt{ipi\_lag}), whereas the semi-structured model provides smaller and more stable estimates, suggesting that potential collinearity and nonlinear interactions among these indicators are being absorbed by the unstructured component.

Importantly, adding macro variables does not substantially alter the main borrower/loan conclusions relative to the no-macro specification. The dominant effects remain stable across both models, i.e.\ \texttt{fico} is protective against delinquency entry ($0\!\to\!1$) and supports cures ($1\!\to\!0$, $2\!\to\!0$), the occupancy indicators remain consistently adverse for cure transitions and also negative for deterioration and default transitions, and \texttt{cov\_dummy} continues to show a pronounced increase in deterioration/default risk. Coefficient magnitudes for \texttt{int\_rt}, \texttt{ltv}, and the DTI categories change only slightly, showing that macro covariates add incremental signal rather than displacing the core borrower/loan risk drivers.

As in the no-macro case, the WOE-transformed high-cardinality terms remain influential. Under the semi-structured model, their linear effects become substantially larger once macro variables are included, which may reflect a reallocation of explanatory power between the structured and unstructured components when additional (and potentially correlated) predictors are introduced.

\begin{figure}[htbp]
	\centering
	  \includegraphics[width=0.85\textwidth]{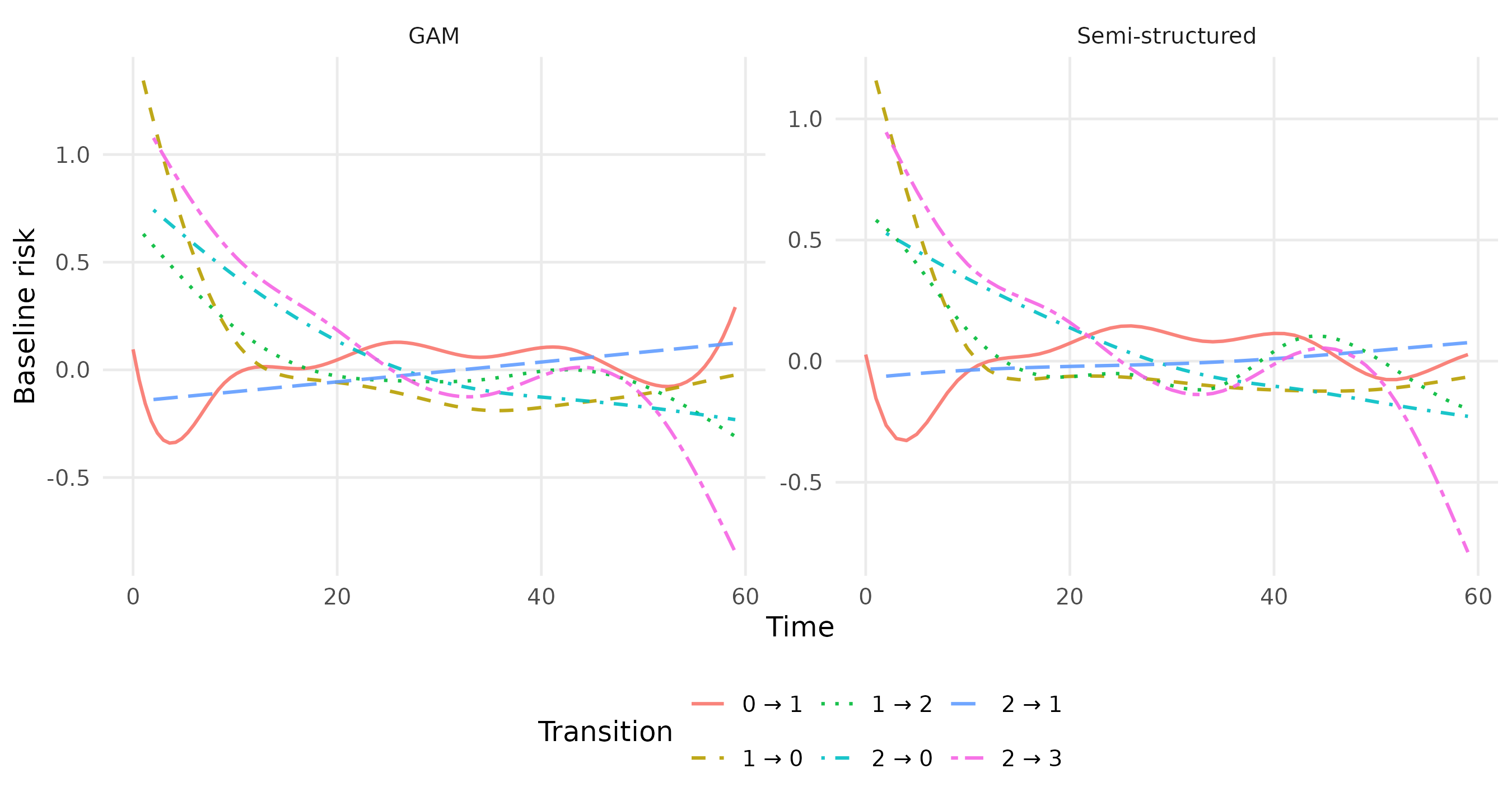}
	  \caption{Estimated time-dependent baseline risk terms $f_{kl}(t)$ (log-odds scale) by transition for the GAM and the semi-structured model, estimated with macroeconomic covariates. Patterns are similar to the no-macro specification, indicating that macro covariates do not materially alter the duration dependence in the transition dynamics.}
  \label{fig:baseline_mac1}
\end{figure}

\paragraph{Baseline risk terms.}
Figure~\ref{fig:baseline_mac1} reports the corresponding baseline risk terms when macro covariates are included. The duration patterns are broadly unchanged relative to Figure~\ref{fig:baseline_mac0}, i.e.\ early transition propensities out of state 1 and into default from state 2 remain pronounced, and $2\!\to\!1$ continues to rise gradually with duration. In other words, macro covariates do not appear to absorb the main duration dependence in the transition process, and the baseline terms continue to capture systematic time-since-origination effects that persist after conditioning on both borrower/loan and macroeconomic information.

We now turn to predictive performance and probability calibration.

\subsection{Performance}\label{subsec:performance}

Throughout, we fix an evaluation window $[t_1,t_2]$ and consider out-of-sample and out-of-time subjects $i=1,\dots,\tilde{N}$ with observed end state $Z_i(t_2)\in\{0,\dots,K-1\}$. Let
\[
\widehat{\mathbf p}_i(t_1,t_2)
=\bigl(\widehat p_{i0}(t_1,t_2),\dots,\widehat p_{i,K-1}(t_1,t_2)\bigr)
\]
denote the model predicted distribution of $Z_i(t_2)$ given $Z_i(t_1)$ and covariates, estimated following \eqref{eq:subj_prob}. We write $Y_{ij}=\mathbbm{1}\{Z_i(t_2)=j\}$ for the one-hot outcome. The metrics below are reported on the test set, however, when a calibration/optimisation step is needed, particularly for cut-points, it is performed on a calibration subset of the training data.

\subsubsection{Metrics}

We measure performance along three dimensions, namely, discrimination, calibration, and accuracy.

\paragraph{Discrimination: ROC AUCs.}
We report two discriminative metrics, (i) a \emph{multiclass} AUC and (ii) a \emph{one-vs-all} AUC version.
\begin{enumerate}
  \item \textbf{Multiclass AUC.} Following \citet{hand2001simple}, the multiclass AUC is defined as the average of pairwise AUCs across all class pairs:
  \[
  \mathrm{AUC}_{\text{multi}}
  \;=\; \frac{2}{K(K-1)} \sum_{0\le r<s\le K-1}
  \frac{\mathrm{AUC}(r\!\mid\!s) + \mathrm{AUC}(s\!\mid\!r)}{2},
  \]
  where $\mathrm{AUC}(r\!\mid\!s)$ is the usual binary AUC computed on the subset $\{i: Z_i(t_2)\in\{r,s\}\}$ with class $r$ treated as positive and $s$ as negative, using score $\widehat p_{ir}(t_1,t_2)$ \citep{fawcett2006introduction}. This measures the model's ability to rank the correct class across all pairs.

  \item \textbf{One-vs-all AUCs.} For each $j\in\{0,\dots,K-1\}$, we treat class $j$ as positive with labels $Y_{ij}$ and use scores $\widehat p_{ij}(t_1,t_2)$. The per-class AUC is
  \[
  \mathrm{AUC}_j = \mathrm{AUC}\!\left(Y_{ij},\, \widehat p_{ij}(t_1,t_2)\right),
  \]
  and we report the prevalence-weighted average $\sum_{j=0}^{K-1} \widehat\omega_j\,\mathrm{AUC}_j$, where $\widehat\omega_j=\frac{1}{\tilde N}\sum_{i=1}^{\tilde N} Y_{ij}$.
\end{enumerate}

\paragraph{Calibration: Brier score and Expected Calibration Error (ECE).}
Calibration evaluates agreement between predicted probabilities and observed frequencies.

\begin{enumerate}
  \item \textbf{Multiclass Brier score.} The mean-squared Brier score \citep{brier1950} generalised to $K$ classes is
  \[
  \mathrm{BS}_{\text{multi}} \;=\; \frac{1}{N K} \sum_{i=1}^N \sum_{j=0}^{K-1} \bigl(\widehat p_{ij}(t_1,t_2) - Y_{ij}\bigr)^2,
  \]
  i.e., the average squared error per class (multiplying by $K$ recovers the conventional multiclass scaling). 

  \item \textbf{Expected Calibration Error (ECE).}
  For each class $j$, we partition the probability range $[0,1]$ into bins $\{I_m\}_{m=1}^M$ and let $B_{jm}=\{i:\widehat p_{ij}(t_1,t_2)\in I_m\}$ with size $n_{jm}=|B_{jm}|$. For bins with $n_{jm}>0$, define the within-bin average score and empirical accuracy
  \[
  \widehat{\mu}_{jm}=\frac{1}{n_{jm}}\sum_{i\in B_{jm}}\widehat p_{ij}(t_1,t_2),
  \qquad
  \widehat a_{jm}=\frac{1}{n_{jm}}\sum_{i\in B_{jm}} Y_{ij}.
  \]
  The class-wise ECE is \citep{{naeini2015obtaining,guo2017calibration}}
  \[
  \mathrm{ECE}_j \;=\; \sum_{m=1}^M \frac{n_{jm}}{N}\,\bigl|\widehat{\mu}_{jm}-\widehat a_{jm}\bigr|.
  \]
  We then obtain the average by $\mathrm{ECE}_{\text{multi}}=\frac{1}{K}\sum_{j=0}^{K-1} \mathrm{ECE}_j$.
\end{enumerate}

\paragraph{Accuracy.}
Following \citet{Djeundje2018}, we summarise end-state prediction via a calibrated cut-point rule applied \emph{conditional on the starting state at $t_1$}. For each start state $k$, let $\widehat{\mathbf p}_i(t_1,t_2)=(\widehat p_{i0}(t_1,t_2),\ldots,\widehat p_{i,K-1}(t_1,t_2))$ be the predicted distribution for subjects with $Z_i(t_1)=k$, and let $\mathbf c_k=(c_{k0},\ldots,c_{k,K-1})\in(0,1)^K$ be a vector of class-specific cut-points (constrained to $c_{kj}\in[\varepsilon,1-\varepsilon]$ for a small $\varepsilon>0$ to avoid numerical instability). Define a discrepancy score $S_{ij}^{(k)}$ by:\footnote{We also considered alternative discrepancy scores, including a raw difference $\widehat p_{ij}(t_1,t_2)-c_{kj}$ and a standardised version $(\widehat p_{ij}(t_1,t_2)-c_{kj})/\widehat\sigma_{kj}$, where $\widehat\sigma_{kj}$ is the empirical standard deviation of $\widehat p_{ij}(t_1,t_2)$ among $\{i:Z_i(t_1)=k\}$. The main conclusions were not sensitive to this choice.}
\[
S_{ij}^{(k)}=\frac{\widehat p_{ij}(t_1,t_2)-c_{kj}}{c_{kj}}.
\]

The predicted class is
\[
\widehat Z_i^{(k)}(t_2) \;=\; \arg\max_{j\in\{0,\dots,K-1\}} S_{ij}^{(k)}.
\]
Cut-points $\mathbf c_k$ are estimated on a calibration set by maximising the proportion of correct classifications among $\{i:Z_i(t_1)=k\}$. We report overall accuracy.

\subsubsection{Results}

Tables~\ref{tab:perf_mac0_h6}--\ref{tab:perf_mac1_h12} compare GAM to the semi-structured model across prediction spans and two horizons ($h=6$ and $h=12$ months), estimated both without and with macroeconomic covariates. Across all settings, differences are small and concentrated in discrimination for the earliest spans. Brier scores are essentially identical at the reported precision.

\paragraph{No macroeconomic variables.}
For $h=6$ (Table~\ref{tab:perf_mac0_h6}), the semi-structured model delivers discrimination gains in the early spans, with MultiAUC improving from 0.741 to 0.753 for span 6--12 (+0.012) and from 0.791 to 0.800 for span 12--18 (+0.009). The gains vanish by span 18--24, where MultiAUC is identical (0.755) and AUC1vsA is virtually unchanged (0.841 vs.\ 0.840). Calibration is slightly weaker for the semi-structured model in these early spans, with ECE increasing from 0.002 to 0.003 (spans 6--12 and 12--18) and from 0.005 to 0.008 (span 18--24).

\begin{table}[t]
\centering
\caption{Performance by span at $h=6$ months (no macro): GAM vs.\ semi-structured.}
\label{tab:perf_mac0_h6}
\centering\begingroup\fontsize{9}{11}\selectfont

\resizebox{\ifdim\width>\linewidth\linewidth\else\width\fi}{!}{
\begin{tabular}{lrrrrrrrrrl}
\toprule
\multicolumn{1}{c}{ } & \multicolumn{5}{c}{GAM} & \multicolumn{5}{c}{Semi-structured} \\
\cmidrule(l{3pt}r{3pt}){2-6} \cmidrule(l{3pt}r{3pt}){7-11}
Span & MultiAUC & AUC1vsA & Brier & ECE & ACC & MultiAUC & AUC1vsA & Brier & ECE & ACC\\
\midrule
6-12 & 0.741 & 0.827 & 0.002 & 0.002 & 0.995 & 0.753 & 0.838 & 0.002 & 0.003 & 0.995\\
12-18 & 0.791 & 0.871 & 0.004 & 0.002 & 0.992 & 0.800 & 0.882 & 0.004 & 0.003 & 0.992\\
18-24 & 0.755 & 0.841 & 0.005 & 0.005 & 0.989 & 0.755 & 0.840 & 0.005 & 0.008 & 0.988\\
\bottomrule
\end{tabular}}
\endgroup{}

\end{table}

For $h=12$ (Table~\ref{tab:perf_mac0_h12}), discrimination differences are smaller. MultiAUC increases from 0.708 to 0.715 in span 6--18 (+0.007), while later spans are effectively tied (e.g., 0.720 vs.\ 0.721 for span 12--24 and 0.727 vs.\ 0.727 for span 18--30). ECE is again marginally higher under the semi-structured model (e.g., 0.003 to 0.005 in spans 6--18 and 12--24, and 0.007 to 0.010 in span 18--30).

\begin{table}[t]
\centering
\caption{Performance by span at $h=12$ months (no macro): GAM vs.\ semi-structured.}
\label{tab:perf_mac0_h12}
\centering\begingroup\fontsize{9}{11}\selectfont

\resizebox{\ifdim\width>\linewidth\linewidth\else\width\fi}{!}{
\begin{tabular}{lrrrrrrrrrl}
\toprule
\multicolumn{1}{c}{ } & \multicolumn{5}{c}{GAM} & \multicolumn{5}{c}{Semi-structured} \\
\cmidrule(l{3pt}r{3pt}){2-6} \cmidrule(l{3pt}r{3pt}){7-11}
Span & MultiAUC & AUC1vsA & Brier & ECE & ACC & MultiAUC & AUC1vsA & Brier & ECE & ACC\\
\midrule
6-18 & 0.708 & 0.778 & 0.002 & 0.003 & 0.995 & 0.715 & 0.783 & 0.002 & 0.005 & 0.995\\
12-24 & 0.720 & 0.780 & 0.004 & 0.003 & 0.993 & 0.721 & 0.778 & 0.004 & 0.005 & 0.993\\
18-30 & 0.727 & 0.797 & 0.005 & 0.007 & 0.988 & 0.727 & 0.796 & 0.005 & 0.010 & 0.988\\
\bottomrule
\end{tabular}}
\endgroup{}

\end{table}

\paragraph{With macroeconomic variables.}
Including macroeconomic indicators does not improve out-of-time performance (Tables~\ref{tab:perf_mac1_h6}--\ref{tab:perf_mac1_h12}). For $h=6$, the semi-structured model no longer consistently dominates, and it is slightly better at span 6--12 (MultiAUC 0.736 to 0.740) but matches the GAM at span 12--18 (0.785 vs.\ 0.785) and underperforms at span 18--24 (0.755 vs.\ 0.745). For $h=12$, the pattern is similar, i.e.\ near the same at span 6--18 (0.705 vs.\ 0.706) and small disadvantages for the semi-structured model at later spans (e.g., 0.718 vs.\ 0.713 for span 12--24 and 0.730 vs.\ 0.723 for span 18--30). Calibration (ECE) is essentially unchanged between models in the macro specification (often identical to three decimals).

\begin{table}[t]
\centering
\caption{Performance by span at $h=6$ months (with macro): GAM vs.\ semi-structured.}
\label{tab:perf_mac1_h6}
\centering\begingroup\fontsize{9}{11}\selectfont

\resizebox{\ifdim\width>\linewidth\linewidth\else\width\fi}{!}{
\begin{tabular}{lrrrrrrrrrl}
\toprule
\multicolumn{1}{c}{ } & \multicolumn{5}{c}{GAM} & \multicolumn{5}{c}{Semi-structured} \\
\cmidrule(l{3pt}r{3pt}){2-6} \cmidrule(l{3pt}r{3pt}){7-11}
Span & MultiAUC & AUC1vsA & Brier & ECE & ACC & MultiAUC & AUC1vsA & Brier & ECE & ACC\\
\midrule
6-12 & 0.736 & 0.822 & 0.002 & 0.002 & 0.995 & 0.740 & 0.825 & 0.002 & 0.002 & 0.995\\
12-18 & 0.785 & 0.864 & 0.004 & 0.002 & 0.992 & 0.785 & 0.861 & 0.004 & 0.002 & 0.992\\
18-24 & 0.755 & 0.843 & 0.005 & 0.006 & 0.989 & 0.745 & 0.829 & 0.005 & 0.006 & 0.989\\
\bottomrule
\end{tabular}}
\endgroup{}

\end{table}

\begin{table}[t]
\centering
\caption{Performance by span at $h=12$ months (with macro): GAM vs.\ semi-structured.}
\label{tab:perf_mac1_h12}
\centering\begingroup\fontsize{9}{11}\selectfont

\resizebox{\ifdim\width>\linewidth\linewidth\else\width\fi}{!}{
\begin{tabular}{lrrrrrrrrrl}
\toprule
\multicolumn{1}{c}{ } & \multicolumn{5}{c}{GAM} & \multicolumn{5}{c}{Semi-structured} \\
\cmidrule(l{3pt}r{3pt}){2-6} \cmidrule(l{3pt}r{3pt}){7-11}
Span & MultiAUC & AUC1vsA & Brier & ECE & ACC & MultiAUC & AUC1vsA & Brier & ECE & ACC\\
\midrule
6-18 & 0.705 & 0.776 & 0.002 & 0.003 & 0.995 & 0.706 & 0.779 & 0.002 & 0.003 & 0.995\\
12-24 & 0.718 & 0.781 & 0.004 & 0.003 & 0.993 & 0.713 & 0.777 & 0.004 & 0.003 & 0.993\\
18-30 & 0.730 & 0.802 & 0.005 & 0.008 & 0.988 & 0.723 & 0.793 & 0.005 & 0.007 & 0.988\\
\bottomrule
\end{tabular}}
\endgroup{}

\end{table}

Overall accuracy remains high (roughly 0.988--0.995) across models, horizons, and spans, reflecting strong class imbalance in multi-state delinquency outcomes (also seen in previous works, e.g.,\ \citet{Djeundje2018}). We therefore treat accuracy as secondary and focus on discrimination and calibration.

\section{Conclusion} \label{sec:concl}
This paper studies mortgage delinquency dynamics with a discrete-time multi-state model. We propose a semi-structured specification that combines an interpretable structured predictor (linear effects and smooth terms) with an unstructured neural component that can learn nonlinearities and interactions. To keep the two parts distinct, we enforce identifiability by orthogonalising the neural representation with respect to the structured design. Although motivated by mortgage delinquency, the modelling framework is generic and applies to any discrete-time multi-state process where one wishes to retain an interpretable core while allowing flexible residual structure.

Moreover, we present exact discrete-time transformations from binary logistic transition models to valid competing transition probabilities. This avoids continuous-time approximations that can add additional noise when data are recorded at discrete intervals, as is typical for monthly mortgage performance panels.

In simulations, the method recovers baseline and covariate effects while using the unstructured component to capture interaction structure. In the empirical study using Freddie Mac mortgage data and out-of-time evaluation, the semi-structured model and GAM agree on the main directional effects. Credit score is a dominant driver of delinquency entry and cures, and duration dependence remains strong even after conditioning on observed covariates. Performance differences are modest: without macro variables, the semi-structured model shows discrimination gains, mainly in the earliest spans, while Brier scores are essentially unchanged and calibration (ECE) is sometimes slightly worse. Adding macroeconomic indicators does not improve out-of-time performance in this setting and does not materially change the estimated baseline duration profiles.

Overall, the results suggest that the semi-structured specification can offer incremental gains in discrimination while keeping key effects interpretable. Importantly, in this application, the added flexibility does not materially worsen overall accuracy or the Brier score relative to the structured benchmark. This makes the semi-structured model a ``low-regret'' extension of a structured additive baseline. That is, when additional nonlinearities or interaction structure are present, they can be captured by the neural component, and when they are not, performance remains close to the structured model. The limited value of macro variables here likely reflects the out-of-time design and the fact that much of the relevant risk signal is already captured by borrower, loan, and performance-history information.

We envision several extensions. From an applied perspective,  mortgages have other exit events (e.g., prepayment and liquidation), and adding these as extra states or competing absorbing risks would broaden the scope. From a methodological perspective, calibration could be improved post hoc by fitting a recalibration map on a held-out calibration set, applied to the predicted end-state probabilities at the evaluation horizon (e.g., temperature scaling or isotonic regression). In addition, uncertainty quantification could be estimated with approaches that respect loan-level dependence (e.g., block bootstrap or Bayesian/ensemble variants). Furthermore, the unstructured component can incorporate non-traditional inputs (e.g., transactional signals, text, or images) when available, while the structured part continues to provide an interpretable summary of the main drivers of transition probabilities. Finally, all methodological advancements could be adjusted and applied to various other domains beyond credit risk. 

\paragraph{Acknowledgements} Victor Medina-Olivares and Nadja Klein acknowledge financial support from the German Research Foundation (DFG) through the Emmy Noether grant KL3037/1-1.

\clearpage

{\LARGE{\begin{center}{\textbf{APPENDIX}}\end{center}}}
\appendix

\section{Summary statistics for test set} \label{app:sum_stat}

\begin{table}[h]
    \centering
    \begin{tabular}{lrrrrr}
        \toprule
            \multicolumn{1}{c}{\textbf{Variable}} & \multicolumn{1}{c}{\textbf{Min.}} & \multicolumn{1}{c}{\textbf{Mean}} & \multicolumn{1}{c}{\textbf{Median}} & \multicolumn{1}{c}{\textbf{Max.}} & \multicolumn{1}{c}{\textbf{SD.}} \\
        \midrule
            \textit{Credit Score} & 479.00 & 744.82 & 752.00 & 832.00 & 46.81 \\
            \textit{Original Interest Rate (\%)} & 2.88 & 4.70 & 4.75 & 6.88 & 0.49 \\
            \textit{Original Loan-to-Value Ratio (\%)} & 6.00 & 75.48 & 80.00 & 135.00 & 17.67 \\
            \textit{Mortgage Insurance (\%)} & 0.00 & 8.12 & 0.00 & 35.00 & 12.16 \\
            \textit{Original Loan Term (months)} & 96.00 & 329.79 & 360.00 & 360.00 & 65.69 \\
            \textit{Original UPB (\$)} & 16,000.00 & 213,876.98 & 188,000.00 & 1,124,000.00 & 117,053.28 \\
        \bottomrule
    \end{tabular}
    \caption{Summary statistics for numerical variables in the test set. More detail on these variables can be found in the \href{https://www.freddiemac.com/research/datasets/sf-loanlevel-dataset}{General User Guide}  for the Freddie Mac Single-Family Loan-Level Dataset. Numerical variables are normalised prior to modelling.}
    \label{tab:sum_stat_num_t}
\end{table}

\begin{table}[h]
    \centering
    \begin{tabular}{l r}
        \toprule
            \textbf{Variable(s)} & \textbf{Cardinality} \\
        \midrule
            \textit{Number of Borrowers}, \textit{First Time Homebuyer}, \textit{Super Conforming} & 2 \\
            \textit{Channel}, \textit{Loan Purpose}, \textit{Mortgage Insurance Cancellation}, \textit{Occupancy Status} & 3 \\
            \textit{Number of Units} & 4 \\
            \textit{Original Debt-to-Income Ratio}, \textit{Property Type} & 5 \\
            \textit{Seller Name} & 32 \\
            \textit{Servicer Name} & 26 \\
            \textit{US state or territory} & 54 \\
        \bottomrule
    \end{tabular}
    \caption{Cardinality of categorical variables in the test set. More detail on these variables can be found in the \href{https://www.freddiemac.com/research/datasets/sf-loanlevel-dataset}{General User Guide} for the Freddie Mac Single-Family Loan-Level Dataset.}
    \label{tab:sum_stat_cat_t}
\end{table}

\section{Data dictionary} \label{app:dict}

\begin{table}[h]
\centering
\begin{adjustbox}{max width=\textwidth}
\begin{tabular}{lll}
\toprule
\textbf{Model term} & \textbf{Freddie Mac field / source} & \textbf{Meaning / level} \\
\midrule
fico & Credit Score & Borrower credit score \\
int\_rt & Original Interest Rate & Note rate at origination \\
ltv & Original LTV & Original loan-to-value ratio \\
orig\_upb & Original UPB & Original unpaid principal balance \\
orig\_loan\_term & Original Loan Term & Original term in months \\
mi\_pct & Mortgage Insurance (\%) & MI percentage at origination \\
cnt\_units & Number of Units & Number of units  \\
cnt\_borrone & Number of Borrowers & Indicator for one borrower (ref: multi-borrower) \\
flag\_fthbY & First-Time Homebuyer Flag & Y = first-time homebuyer (ref: N) \\
flag\_scY & Super Conforming Flag & Y = super conforming (ref: N) \\
channelC & Channel & C = Correspondent (ref: B = Broker) \\
channelR & Channel & R = Retail (ref: B)\\
loan\_purposeN & Loan Purpose & N = No cash-out refinance (ref: C = Cash-out refinance) \\
loan\_purposeP & Loan Purpose & P = Purchase (ref: C) \\

occpy\_stsP & Occupancy Status & P = Principal residence (ref: I = Investment property) \\
occpy\_stsS & Occupancy Status & S = Second home (ref: I)\\

prop\_typeCP & Property Type & CP = Co-op (ref: CO = Condominium) \\
prop\_typeMH & Property Type & MH = Manufactured housing (ref: CO) \\
prop\_typePU & Property Type & PU = Planned unit development (ref: CO) \\
prop\_typeSF & Property Type & SF = Single-family (ref: CO) \\
mi\_cxl\_indN & MI Cancellation Indicator & N = MI not cancelled (ref: 7 = Not applicable) \\
mi\_cxl\_indY & MI Cancellation Indicator & Y = MI cancelled (ref: 7) \\
dti1 & Original DTI & DTI $\le 20\%$ (ref: dti4 = $40\%<\text{DTI}\le 65\%$) \\
dti2 & Original DTI & $20\%<\text{DTI}\le 30\%$ (ref: dti4) \\
dti3 & Original DTI & $30\%<\text{DTI}\le 40\%$ (ref: dti4) \\
dti5 & Original DTI & $\text{DTI}>65\%$ or not available (ref: dti4) \\
us\_state\_te & Property State / Territory & High-cardinality state/territory, WOE-transformed \\
seller\_name\_te & Seller Name & High-cardinality seller, WOE-transformed \\
servicer\_name\_te & Servicer Name & High-cardinality servicer, WOE-transformed \\
cov\_dummy & Constructed for Mar-Dec 2020 & COVID/pandemic-period flag \\
cci\_lag & Macroeconomic & Consumer confidence index \\
cpi\_lag & Macroeconomic & Consumer price index / inflation proxy  \\
ipi\_lag & Macroeconomic & Industrial production index/ real activity proxy \\
mort30\_lag & Macroeconomic & 30-year fixed mortgage rate \\
sp500\_lag & Macroeconomic & S\&P 500 index  \\
unrt\_lag & Macroeconomic & Unemployment rate  \\
\bottomrule
\end{tabular}
\end{adjustbox}
\caption{Mapping from model coefficient terms to Freddie Mac variables, category codes, and macroeconomic series.}
\label{tab:covariate_dictionary}
\end{table}

\section{Full linear-effect tables with macroeconomic variables (GAM and semi-structured)} \label{app:lin_mac1}

This appendix reports the full tables of estimated \emph{linear} effects for the specification that augments borrower- and loan-level covariates with lagged macroeconomic indicators. The tables are discussed in Section~\ref{subsubsec:mac1}, but are shown here in full due to their size. Entries are reported on the log-odds scale for each permissible transition.

\begin{landscape}
  \begin{table}[p]
  \centering
  \caption{Linear covariate effects (mean and 95\% interval). GAM with macroeconomic variables.}
  \label{tab:lin_eff_gam_mac1}
  \begin{tabular}{lcccccc}
    \toprule
    \textbf{Name} & 0 $\to$ 1 & 1 $\to$ 0 & 1 $\to$ 2 & 2 $\to$ 0 & 2 $\to$ 1 & 2 $\to$ 3 \\
    \midrule
    cci\_lag & -1.19 (-1.57, -0.81) & -1.27 (-1.98, -0.56) & -1.88 (-2.87, -0.89) & -0.65 (-2.76, 1.47) & 0.14 (-2.34, 2.63) & -5.05 (-6.89, -3.21) \\
    channelC & -0.12 (-0.17, -0.07) & 0.05 (-0.05, 0.15) & 0.08 (-0.05, 0.21) & -0.26 (-0.54, 0.01) & 0.21 (-0.14, 0.57) & -0.30 (-0.52, -0.08) \\
    channelR & -0.27 (-0.32, -0.22) & 0.09 (-0.01, 0.18) & 0.24 (0.12, 0.36) & -0.14 (-0.40, 0.11) & 0.09 (-0.25, 0.43) & -0.21 (-0.41, -0.01) \\
    cnt\_borrone & 0.48 (0.45, 0.51) & -0.19 (-0.25, -0.12) & 0.05 (-0.02, 0.13) & -0.03 (-0.20, 0.13) & 0.03 (-0.17, 0.24) & -0.02 (-0.15, 0.10) \\
    cnt\_units & -0.04 (-0.10, 0.02) & 0.17 (0.02, 0.32) & 0.15 (-0.03, 0.33) & 0.03 (-0.33, 0.40) & -0.12 (-0.63, 0.38) & 0.05 (-0.22, 0.32) \\
    cov\_dummy & 1.10 (1.05, 1.15) & 0.01 (-0.08, 0.10) & 1.43 (1.32, 1.54) & 0.35 (0.11, 0.59) & -0.36 (-0.65, -0.08) & 1.34 (1.14, 1.54) \\
    cpi\_lag & -0.29 (-0.38, -0.21) & -0.08 (-0.24, 0.07) & -0.39 (-0.60, -0.17) & 0.52 (0.07, 0.98) & 0.27 (-0.25, 0.78) & -0.07 (-0.48, 0.35) \\
    dti1 & 0.15 (0.07, 0.23) & -0.16 (-0.33, 0.01) & -0.06 (-0.29, 0.16) & -0.36 (-0.85, 0.14) & -0.27 (-0.88, 0.33) & -0.17 (-0.59, 0.24) \\
    dti2 & 0.34 (0.26, 0.41) & -0.20 (-0.36, -0.04) & 0.01 (-0.20, 0.23) & -0.38 (-0.85, 0.09) & -0.27 (-0.85, 0.30) & -0.22 (-0.62, 0.17) \\
    dti3 & 0.48 (0.41, 0.56) & -0.39 (-0.55, -0.23) & -0.05 (-0.26, 0.17) & -0.28 (-0.75, 0.18) & -0.38 (-0.95, 0.19) & -0.18 (-0.57, 0.21) \\
    dti5 & 0.42 (0.32, 0.51) & -0.32 (-0.51, -0.13) & -0.12 (-0.37, 0.13) & -0.46 (-1.02, 0.10) & -0.96 (-1.67, -0.26) & -0.05 (-0.51, 0.41) \\
    fico & -5.73 (-5.87, -5.59) & 2.39 (2.16, 2.63) & 1.39 (1.09, 1.69) & 1.51 (0.88, 2.14) & -0.91 (-1.75, -0.07) & 1.47 (0.98, 1.97) \\
    flag\_fthbY & -0.09 (-0.13, -0.05) & -0.03 (-0.11, 0.06) & 0.12 (0.02, 0.22) & 0.13 (-0.09, 0.34) & 0.13 (-0.14, 0.40) & 0.05 (-0.12, 0.22) \\
    flag\_scY & 0.04 (-0.05, 0.13) & 0.10 (-0.08, 0.28) & -0.06 (-0.28, 0.16) & -0.16 (-0.65, 0.34) & -0.56 (-1.27, 0.14) & -0.30 (-0.66, 0.05) \\
    int\_rt & 1.18 (1.03, 1.34) & -0.33 (-0.65, -0.02) & -0.11 (-0.50, 0.28) & -0.91 (-1.60, -0.22) & 0.05 (-0.82, 0.92) & -0.59 (-1.13, -0.04) \\
    ipi\_lag & -0.10 (-1.07, 0.87) & 6.16 (4.31, 8.01) & -7.85 (-10.41, -5.30) & 6.42 (0.84, 12.01) & 7.33 (1.02, 13.63) & -1.75 (-6.36, 2.85) \\
    loan\_purposeN & -0.07 (-0.12, -0.02) & 0.03 (-0.06, 0.13) & -0.05 (-0.17, 0.07) & 0.09 (-0.16, 0.35) & 0.23 (-0.10, 0.56) & -0.17 (-0.37, 0.04) \\
    loan\_purposeP & -0.03 (-0.08, 0.01) & 0.13 (0.04, 0.22) & -0.12 (-0.24, -0.01) & 0.11 (-0.14, 0.36) & 0.13 (-0.21, 0.46) & -0.24 (-0.44, -0.04) \\
    ltv & 0.61 (0.21, 1.02) & -0.91 (-1.71, -0.11) & 1.66 (0.64, 2.68) & -1.45 (-3.75, 0.85) & -2.05 (-5.11, 1.00) & 0.40 (-1.35, 2.14) \\
    mi\_cxl\_indN & 0.28 (0.16, 0.40) & -0.37 (-0.60, -0.13) & 0.03 (-0.26, 0.32) & 0.48 (-0.15, 1.11) & 0.93 (0.20, 1.67) & 0.48 (-0.01, 0.97) \\
    mi\_cxl\_indY & -0.01 (-0.12, 0.10) & -0.19 (-0.39, 0.02) & -0.07 (-0.32, 0.18) & 0.14 (-0.43, 0.71) & 0.72 (0.06, 1.38) & -0.05 (-0.50, 0.39) \\
    mi\_pct & -0.03 (-0.21, 0.14) & 0.03 (-0.27, 0.34) & -0.25 (-0.62, 0.12) & -0.78 (-1.59, 0.03) & -0.94 (-1.89, 0.01) & -0.58 (-1.21, 0.05) \\
    mort30\_lag & 1.72 (1.19, 2.25) & 1.39 (0.40, 2.38) & 2.14 (0.74, 3.55) & -0.02 (-2.91, 2.87) & -2.11 (-5.46, 1.24) & 3.90 (1.34, 6.46) \\
    occpy\_stsP & -0.05 (-0.11, 0.01) & -0.19 (-0.32, -0.07) & -0.20 (-0.36, -0.04) & -0.75 (-1.10, -0.40) & -0.57 (-1.06, -0.09) & -0.58 (-0.88, -0.29) \\
    occpy\_stsS & -0.15 (-0.26, -0.04) & -0.30 (-0.53, -0.08) & -0.36 (-0.65, -0.08) & -0.79 (-1.41, -0.16) & -0.41 (-1.27, 0.45) & -0.42 (-0.93, 0.08) \\
    orig\_loan\_term & -0.03 (-0.11, 0.04) & -0.19 (-0.33, -0.04) & 0.03 (-0.16, 0.21) & 0.17 (-0.21, 0.54) & -0.22 (-0.70, 0.25) & 0.25 (-0.05, 0.55) \\
    orig\_upb & 0.79 (0.65, 0.94) & -0.23 (-0.52, 0.06) & 0.44 (0.08, 0.80) & -0.27 (-1.01, 0.47) & 0.71 (-0.21, 1.63) & 0.44 (-0.13, 1.01) \\
    prop\_typeCP & -0.38 (-0.88, 0.11) & -0.03 (-1.24, 1.18) & 1.28 (0.02, 2.55) & -0.81 (-2.26, 0.64) & -0.37 (-2.03, 1.30) & -1.12 (-2.55, 0.31) \\
    prop\_typeMH & 0.20 (-0.02, 0.41) & 0.49 (0.01, 0.97) & 0.29 (-0.31, 0.89) & -0.23 (-1.40, 0.94) & -0.34 (-2.14, 1.46) & -0.36 (-1.30, 0.59) \\
    prop\_typePU & 0.10 (0.04, 0.16) & -0.04 (-0.17, 0.09) & -0.08 (-0.23, 0.08) & -0.08 (-0.42, 0.25) & -0.14 (-0.60, 0.31) & -0.10 (-0.37, 0.16) \\
    prop\_typeSF & 0.17 (0.11, 0.22) & -0.13 (-0.25, -0.01) & -0.23 (-0.37, -0.08) & -0.12 (-0.43, 0.19) & -0.11 (-0.54, 0.32) & -0.30 (-0.55, -0.06) \\
    seller\_name\_te & 2.45 (0.12, 4.79) & 2.31 (1.68, 2.94) & 2.84 (2.19, 3.49) & 2.32 (1.13, 3.51) & 4.95 (3.12, 6.78) & 3.44 (2.13, 4.75) \\
    servicer\_name\_te & 5.38 (3.07, 7.69) & 1.62 (0.95, 2.29) & 0.76 (0.01, 1.51) & 3.35 (2.43, 4.27) & 1.38 (-1.16, 3.91) & 1.94 (1.00, 2.88) \\
    sp500\_lag & 0.37 (0.24, 0.51) & -0.03 (-0.30, 0.23) & 0.73 (0.38, 1.08) & 0.20 (-0.56, 0.95) & -0.54 (-1.45, 0.37) & 1.21 (0.56, 1.86) \\
    unrt\_lag & -0.39 (-0.48, -0.31) & 0.30 (0.14, 0.46) & -0.99 (-1.20, -0.77) & 0.51 (0.05, 0.97) & 0.56 (0.02, 1.09) & -0.59 (-0.97, -0.21) \\
    us\_state\_te & 3.50 (1.17, 5.83) & 3.46 (2.91, 4.01) & 3.47 (2.95, 3.99) & 4.22 (3.43, 5.00) & 4.90 (3.65, 6.15) & 4.33 (3.39, 5.27) \\
    \bottomrule
  \end{tabular}%
\end{table}

\end{landscape}

\begin{landscape}
  \begin{table}[p]
  \centering
  \caption{Linear covariate effects (mean and 95\% interval). Semi-structured with macroeconomic variables.}
  \label{tab:lin_eff_deep_mac1}
  \begin{tabular}{lcccccc}
    \toprule
    \textbf{Name} & 0 $\to$ 1 & 1 $\to$ 0 & 1 $\to$ 2 & 2 $\to$ 0 & 2 $\to$ 1 & 2 $\to$ 3 \\
    \midrule
    cci\_lag & -0.74 (-1.54, 0.16) & -1.01 (-1.76, -0.27) & -1.31 (-2.52, -0.04) & -0.94 (-2.16, 0.49) & -0.53 (-1.71, 0.83) & -2.06 (-3.12, -0.65) \\
    channelC & -0.16 (-0.20, -0.11) & 0.01 (-0.13, 0.14) & 0.04 (-0.08, 0.17) & -0.18 (-0.44, 0.04) & 0.23 (-0.06, 0.55) & -0.24 (-0.45, -0.04) \\
    channelR & -0.29 (-0.33, -0.25) & 0.07 (-0.06, 0.17) & 0.20 (0.08, 0.33) & -0.11 (-0.33, 0.14) & 0.10 (-0.19, 0.41) & -0.15 (-0.35, 0.03) \\
    cnt\_borrone & 0.47 (0.45, 0.49) & -0.24 (-0.31, -0.16) & 0.03 (-0.04, 0.13) & -0.03 (-0.22, 0.14) & 0.04 (-0.21, 0.21) & -0.00 (-0.13, 0.11) \\
    cnt\_units & -0.07 (-0.13, -0.02) & 0.12 (-0.01, 0.28) & 0.13 (-0.04, 0.30) & -0.04 (-0.50, 0.27) & -0.13 (-0.68, 0.30) & -0.02 (-0.19, 0.22) \\
    cov\_dummy & 1.02 (0.95, 1.10) & -0.05 (-0.17, 0.07) & 1.47 (1.36, 1.59) & 0.21 (-0.00, 0.40) & -0.41 (-0.65, -0.18) & 1.17 (0.96, 1.34) \\
    cpi\_lag & -0.29 (-0.47, -0.13) & -0.13 (-0.30, 0.05) & -0.12 (-0.38, 0.18) & 0.43 (0.05, 0.91) & -0.07 (-0.48, 0.24) & 0.16 (-0.19, 0.58) \\
    dti1 & -0.06 (-0.12, 0.00) & -0.21 (-0.43, 0.02) & -0.09 (-0.35, 0.11) & -0.29 (-0.73, 0.08) & -0.23 (-0.80, 0.42) & -0.19 (-0.63, 0.13) \\
    dti2 & 0.14 (0.07, 0.20) & -0.24 (-0.43, -0.06) & 0.07 (-0.18, 0.29) & -0.32 (-0.72, -0.00) & -0.28 (-0.80, 0.27) & -0.31 (-0.73, 0.01) \\
    dti3 & 0.28 (0.23, 0.35) & -0.43 (-0.61, -0.24) & -0.15 (-0.37, 0.06) & -0.25 (-0.64, 0.10) & -0.35 (-0.88, 0.19) & -0.29 (-0.69, 0.04) \\
    dti5 & 0.17 (0.08, 0.24) & -0.45 (-0.66, -0.23) & -0.14 (-0.39, 0.11) & -0.47 (-0.88, -0.02) & -0.93 (-1.55, -0.29) & -0.20 (-0.70, 0.18) \\
    fico & -5.69 (-5.83, -5.53) & 2.32 (2.13, 2.57) & 1.40 (1.13, 1.70) & 1.27 (0.67, 1.95) & -0.94 (-1.63, -0.30) & 1.48 (1.03, 1.88) \\
    flag\_fthbY & -0.11 (-0.15, -0.07) & -0.02 (-0.14, 0.08) & 0.10 (-0.00, 0.22) & 0.11 (-0.11, 0.32) & 0.13 (-0.13, 0.39) & 0.03 (-0.13, 0.17) \\
    flag\_scY & -0.07 (-0.16, 0.02) & 0.14 (-0.07, 0.31) & -0.10 (-0.28, 0.08) & -0.20 (-0.61, 0.15) & -0.56 (-1.37, 0.20) & -0.31 (-0.64, 0.02) \\
    int\_rt & 1.25 (1.10, 1.39) & -0.42 (-0.70, -0.10) & -0.07 (-0.57, 0.33) & -0.72 (-1.35, -0.04) & -0.07 (-0.94, 0.71) & -0.65 (-1.19, -0.14) \\
    ipi\_lag & -1.27 (-2.65, 0.06) & -1.13 (-2.52, 0.18) & -1.81 (-3.20, -0.49) & -1.46 (-2.90, -0.11) & -1.12 (-2.65, 0.33) & -1.62 (-3.03, -0.25) \\
    loan\_purposeN & -0.07 (-0.12, -0.02) & 0.01 (-0.11, 0.12) & -0.05 (-0.19, 0.07) & 0.09 (-0.18, 0.35) & 0.18 (-0.09, 0.43) & -0.20 (-0.41, 0.00) \\
    loan\_purposeP & -0.03 (-0.08, 0.02) & 0.16 (0.06, 0.26) & -0.11 (-0.23, 0.01) & 0.11 (-0.12, 0.33) & 0.10 (-0.18, 0.50) & -0.27 (-0.48, -0.07) \\
    ltv & 0.68 (0.24, 1.16) & -1.07 (-1.75, -0.29) & 0.70 (-0.17, 1.66) & -0.90 (-2.32, 0.46) & -1.16 (-3.08, 0.56) & 0.04 (-1.33, 1.26) \\
    mi\_cxl\_indN & 0.12 (-0.01, 0.28) & -0.35 (-0.62, -0.08) & 0.06 (-0.17, 0.37) & 0.44 (-0.11, 1.00) & 0.92 (0.31, 1.58) & 0.51 (0.05, 1.01) \\
    mi\_cxl\_indY & -0.19 (-0.32, -0.05) & -0.16 (-0.43, 0.10) & -0.05 (-0.27, 0.19) & 0.09 (-0.39, 0.58) & 0.71 (0.15, 1.44) & -0.03 (-0.36, 0.39) \\
    mi\_pct & 0.18 (-0.04, 0.40) & 0.08 (-0.25, 0.39) & -0.29 (-0.68, 0.03) & -0.80 (-1.40, -0.06) & -1.00 (-1.76, -0.26) & -0.67 (-1.18, -0.09) \\
    mort30\_lag & 0.82 (-0.22, 2.16) & 1.06 (-0.00, 2.30) & -0.11 (-1.45, 1.53) & 0.07 (-1.35, 1.65) & 0.22 (-1.20, 1.96) & -0.18 (-1.59, 1.47) \\
    occpy\_stsP & -0.06 (-0.11, -0.01) & -0.26 (-0.38, -0.13) & -0.24 (-0.41, -0.08) & -0.74 (-1.12, -0.43) & -0.56 (-0.99, -0.14) & -0.67 (-1.02, -0.42) \\
    occpy\_stsS & -0.33 (-0.44, -0.22) & -0.32 (-0.56, -0.08) & -0.31 (-0.62, -0.07) & -0.75 (-1.49, -0.18) & -0.36 (-1.24, 0.57) & -0.55 (-1.00, -0.04) \\
    orig\_loan\_term & -0.06 (-0.13, 0.01) & -0.18 (-0.31, -0.04) & 0.01 (-0.17, 0.23) & 0.02 (-0.28, 0.42) & -0.23 (-0.64, 0.29) & 0.21 (-0.06, 0.48) \\
    orig\_upb & 0.84 (0.69, 1.00) & -0.16 (-0.42, 0.09) & 0.49 (0.10, 0.82) & -0.06 (-0.79, 0.57) & 0.68 (-0.36, 1.67) & 0.43 (-0.03, 0.95) \\
    prop\_typeCP & -7.24 (-8.70, -5.61) & -0.02 (-1.40, 1.28) & 1.31 (-0.08, 2.78) & -0.86 (-2.66, 0.25) & -0.27 (-6.31, 2.00) & -1.28 (-6.98, -0.05) \\
    prop\_typeMH & -0.94 (-1.52, -0.56) & 0.38 (-0.09, 0.91) & 0.30 (-0.20, 0.94) & -0.33 (-2.06, 1.31) & -0.25 (-6.22, 2.06) & -0.35 (-1.64, 1.24) \\
    prop\_typePU & -0.01 (-0.06, 0.04) & -0.07 (-0.23, 0.09) & -0.10 (-0.28, 0.05) & -0.19 (-0.56, 0.12) & -0.12 (-0.52, 0.27) & -0.20 (-0.42, 0.07) \\
    prop\_typeSF & 0.06 (0.02, 0.10) & -0.17 (-0.30, -0.05) & -0.28 (-0.43, -0.16) & -0.21 (-0.55, 0.05) & -0.07 (-0.47, 0.28) & -0.32 (-0.56, -0.05) \\
    seller\_name\_te & 5.25 (3.39, 7.24) & 1.70 (0.92, 2.40) & 1.96 (1.30, 2.72) & 1.30 (0.40, 2.39) & 4.08 (2.45, 5.30) & 1.36 (0.18, 2.36) \\
    servicer\_name\_te & 12.98 (11.06, 15.07) & 1.77 (1.06, 2.51) & 1.26 (0.43, 2.07) & 2.96 (2.08, 3.82) & 2.09 (0.42, 3.99) & 2.04 (1.29, 2.80) \\
    sp500\_lag & 0.34 (0.08, 0.66) & -0.09 (-0.41, 0.19) & 0.17 (-0.30, 0.64) & 0.26 (-0.47, 0.99) & 0.07 (-0.45, 0.60) & 0.57 (0.07, 1.10) \\
    unrt\_lag & -0.46 (-0.58, -0.32) & -0.24 (-0.39, -0.09) & -0.51 (-0.68, -0.36) & -0.14 (-0.40, 0.11) & -0.14 (-0.39, 0.14) & -0.36 (-0.50, -0.16) \\
    us\_state\_te & 9.27 (7.43, 11.17) & 2.83 (2.27, 3.27) & 2.99 (2.48, 3.50) & 3.58 (2.88, 4.28) & 4.78 (3.55, 5.81) & 3.06 (2.25, 3.89) \\
    \bottomrule
  \end{tabular}%
\end{table}

\end{landscape}

\bibliographystyle{apalike}
\bibliography{references}

\end{document}